\input harvmac

\catcode`\@=11

\message{<< Additional Active Characters >>}
\catcode`\|= \active 
\def|{\ifmmode \vert\else \char`\|\fi} 
\def\q@m{\string"}
\catcode`\"=\active \def"{\char`\"}
\toksdef\toks@@=2
\toks@{\do \"}\toks@@ =\expandafter{\dospecials}\xdef\dospecials{\the\toks@@\the\toks@}
\toks@{\do \|}\toks@@ =\expandafter{\dospecials}\xdef\dospecials{\the\toks@@\the\toks@}

%
%

\newcount\a@lignstate  \a@lignstate=0  
\def\hssf{\hskip 0pt plus 1fill minus 1fill}
\def\n@ewaligndefs{\def\center##1{\hssf ##1\hssf\null}
          \def\left##1{##1\hssf\null}
          \def\right##1{\hssf ##1\null}}
\newdimen\trulesize
\let\tr=\trulesize
\trulesize = .4pt
\def\zerocenteredbox#1{\ifmmode \ifinner \setbox2 =\hbox{$#1$}\else 
                           \setbox2 =\hbox{$\displaystyle#1$}\fi
                     \else \setbox2 =\hbox{#1}\fi
      \setbox0=\hbox{\lower.5ex\hbox{$\vcenter{\box2}$}}\ht0=0pt\dp0 =0pt\box0}

\def\modifystrut#1#2#3{\setbox4=\hbox{#1}\dimen0=\ht4
             \advance \dimen0 by #2 \dimen2 = \dp4 
             \advance  \dimen2 by #3 
            \vrule width 0pt height \dimen0 depth \dimen2}


%
\newskip\tcs  
\newtoks\tablespread
\newskip\midtabglue \midtabglue = 0pt plus 1fill
\newtoks\everytable  \everytable = {\relax}

{\catcode`\|=\active   \catcode`\" = \active
  \gdef\begintable{\vbox\bgroup \tcs=.5em 
                  \catcode`\|=\active
                  \catcode`\"=\active
                   \def\:{\relax \vrule height 2.5ex depth .9ex width 0pt}
          \def\-{\ifcase\a@lignstate \fulltablerule{\tr}
                  \else 
                  \thrule{\tr}\fi}
          \let\t@xx =\relax 
          \everycr={\noalign{\global\a@lignstate=0}}
          \def\fulltablerule##1{\noalign{\hrule height
                   ##1}}
          \def\thrule##1{\omit\leaders\hrule height ##1\hfill}
          \def\center{\hskip\tcs\hss ########\hss\hskip\tcs}
          \def\left{\hskip\tcs ########\hss\hskip\tcs}
          \def\right{\hskip\tcs\hss ########\hskip\tcs}
          \def\sprule{\tvrule{2.5\tr}}
          \def|{\ifcase\a@lignstate \def\t@xx{\tvrule{\tr}}\or
                            \def\t@xx{\tvrule{\tr}}\or
                             \def\t@xx{\unskip&\tvrule{\tr}&}\else 
                             \def\t@xx{\tvrule{\tr}}\fi\t@xx}
          \def\|{\ifcase\a@lignstate \def\t@xx{\sprule}\or
                            \def\t@xx{\sprule}\or
                             \def\t@xx{\unskip&\sprule&}\else 
                              \def\t@xx{\sprule}\fi\t@xx}
          \def"{&########&} 
          \def\br##1{\global\a@lignstate=1 ##1\unskip\global\a@lignstate=2&}
          \def\er##1{\global\a@lignstate=3\unskip&##1\unskip
                      \global\a@lignstate=0\cr}
          \def\tvrule##1{\hss\vrule width ##1\hss}
          \def~{\penalty\@M \hphantom{0}}
          \tablespread = {}
          \the\everytable 
               }
  \gdef\begintableformat #1\endtableformat{\offinterlineskip \tabskip = 0pt
       \edef\t@blform{####\tabskip =\midtabglue &#1&####\tabskip=0pt\cr}
                                  \n@ewaligndefs
                                 \def"{\ifcase\a@lignstate \def\t@xx{\relax}\or
                                          \def\t@xx{\relax}\or
                                        \def\t@xx{\unskip&&}\else 
                                        \def\t@xx{\relax}\fi\t@xx}
                                \edef\h@align{\halign \the\tablespread}
                                  \h@align\bgroup\span\t@blform}
      }

\def\use#1{\omit\mscount=#1 \advance\mscount by -1\multiply\mscount by2
                \loop\ifnum\mscount>1 \sp@n\repeat
                \ifnum\mscount>0 \span \else \relax \fi}

\def\sa#1{\setbox0=\hbox{#1}\hbox to \wd0{}}
\def\endtable{\crcr\egroup\egroup}

\catcode`\@=12


\input epsf 

\vskip 1cm

 \Title{ \vbox{\baselineskip12pt\hbox{  Brown Het-1309 }}}
 {\vbox{
\centerline{ Higher dimensional geometries     }
\vskip.08in
\centerline{  related to    }
\vskip.08in
\centerline{  Fuzzy odd-dimensional spheres } }}

\centerline{$\quad$ {  Sanjaye Ramgoolam   } }
\smallskip
\centerline{{\sl Department of Physics}}
\centerline{{\sl Brown  University}}
\centerline{{\sl Providence, RI 02912 }}
\smallskip
\centerline{{ \tt ramgosk@het.brown.edu }}

\vskip .3in 

 We study $SO(m)$ covariant 
 Matrix realizations of $ \sum_{i=1}^{m} X_i^2 = 1 $
 for even $m$ as candidate fuzzy odd spheres following 
 hep-th/0101001. As for the fuzzy four sphere, 
 these Matrix algebras contain more degrees of freedom 
 than the sphere itself and the full set of variables 
 has a geometrical description in terms of a higher dimensional
 coset. The fuzzy $S^{2k-1}  $ is related 
 to  a higher dimensional coset $ { SO(2k) \over U(1) \times U(k-1)
 }$.  These cosets are bundles  where base and fibre are  
hermitian symmetric spaces. 
 The detailed form of the generators and relations for the 
Matrix algebras related to the  fuzzy 
three-spheres suggests Matrix actions which admit the 
fuzzy  spheres as  solutions. These Matrix actions are compared with 
the BFSS, IKKT and BMN Matrix models as well as some others. 
The geometry and combinatorics of fuzzy odd spheres
lead to some remarks on the transverse five-brane problem 
of Matrix theories and the exotic scaling of the entropy of 5-branes 
with the brane number.

\Date{ June   2002 } 


\lref\sphdiv{S. Ramgoolam, 
 `` On spherical harmonics for fuzzy spheres in diverse
dimensions, ''hep-th/0105006, 
 Nucl.Phys. B610 (2001) 461-488  } 
\lref\hdim{ P.M. Ho, S. Ramgoolam, 
``Higher dimensional geometries from matrix brane constructions,''
 hepth/0111278, Nucl.Phys. B627 (2002) 266-288    } 
\lref\dolwit{ L. Dolan and E. Witten, hepth/9910205 }  
\lref\kata{D. Kabat and W. Taylor, 
``Spherical membranes in Matrix theory,'' hep-th/9711078,
Adv.Theor.Math.Phys. 2 (1998) 181-206   } 
\lref\myersdiel{ R. Myers, ``Dielectric-Branes,'' hep-th/9910053  } 
\lref\myrsnabint{ N. Constable, R. Myers, O. Tafjord, ``Non-abelian 
Brane intersections, '' hep-th/0102080, JHEP 0106 (2001) 023   } 
\lref\guram{ Z. Guralnik, S. Ramgoolam
``On the Polarization of Unstable D0-Branes
 into Non-Commutative Odd Spheres,'' hep-th/0101001,   JHEP 0102 (2001) 032 } 
\lref\bmn{ D. Berenstein, J. Maldacena, H. Nastase
``Strings in flat space and pp waves from ${\cal N}=4$ 
Super Yang Mills,''hep-th/0202021, JHEP 0204 (2002) 013  } 
\lref\malstrom{J. Maldacena, A. Strominger,
``AdS3 Black Holes and a Stringy Exclusion Principle,'' hep-th/980408,
  JHEP 9812 (1998) 005  } 
\lref\jevram{A. Jevicki, S. Ramgoolam, 
``Non commutative gravity from the ADS/CFT correspondence,'' 
 hep-th/9902059, JHEP 9904 (1999) 032} 
\lref\holi{P.M. Ho, M. Li,
``Fuzzy Spheres in AdS/CFT Correspondence and 
Holography from Noncommutativity,''  hep-th/0004072, 
Nucl.Phys.B596:259-272,2001  } 
\lref\berkverl{M. Berkooz, H. Verlinde 
``Matrix Theory, AdS/CFT and Higgs-Coulomb Equivalence,'' 
 hep-th/9907100, JHEP 9911 (1999) 037 }  
\lref\kimura{ Y. Kimura ``Noncommutative Gauge Theory on Fuzzy Four-Sphere
 and Matrix Model,''  hep-th/0204256 } 
\lref\bilal{ M. Bagnoud,   L. Carlevaro, A. Bilal, 
  ``Supermatrix models for M-theory based on osp(1|32,R),'' 
    hep-th/0201183 } 
\lref\smol{ L. Smolin, ``M theory as a matrix extension of Chern
 Simons theory,'' hepth/0002009, Nucl.Phys.B591:227-242,2000 } 
\lref\chaudhuri{ S. Chaudhuri, ``Bosonic Matrix Theory and D-branes,'' 
hepth/0205306 } 
\lref\hosph{ P.M. Ho, ``Fuzzy sphere from Matrix model,'' 
hepth/0110165, JHEP 0012 (2000) 015 } 
\lref\iktw{ S.Iso, Y.Kimura, K.Tanaka, K. Wakatsuki, 
 ``Noncommutative Gauge Theory on Fuzzy Sphere from Matrix Model,''
      hep-th/0101102 } 
\lref\ikkt{ N. Ishibashi, H. Kawai, Y. Kitazawa, A. Tsuchiya, 
`` A large-N reduced model as Superstring, '' Nucl. Phys. B498 
(1997)  467-491 } 
\lref\vipul{V. Periwal, ``Matrices on a point as the theory of everything,'' 
Phys.Rev. D55 (1997) 1711-1713 } 
\lref\atwit{ M. Atiyah and E. Witten, 
``M-Theory Dynamics On A Manifold Of $G_2$ Holonomy,'' hep-th/0107177  } 
\lref\guklpe{ S. Gubser, I. Klebanov, A. Peet, 
``Entropy and Temperature of Black 3-Branes,'' 
hep-th/9602135, Phys.Rev. D54 (1996) 3915-3919 } 
\lref\clt{ J.Castelino, S. Lee and W. Taylor IV, ``Longitudinal Five-Branes
as Four Spheres in Matrix Theory,'' Nucl.Phys.{\bf B526} (1998) 334, 
hep-th/9712105.}
\lref\bl{ I. Bandos, J. Lukierski, 
``New superparticle models outside the HLS suersymmetry scheme,''
hep-th/9812074 } 
\lref\fab{ M. Fabinger, ``Higher-Dimensional 
Quantum Hall Effect in String Theory,'' hep-th/0201016, 
 JHEP 0205 (2002) 037 } 
\lref\zhhu{ S.C Zhang, J. Hu 
``A Four Dimensional Generalization of the Quantum Hall Effect,''
cond-mat/0110572,  Science 294 (2001) 823 } 
\lref\sastr{ A. Salam, J. Strathdee, ``On Kaluza-Klein Theory,'' 
Ann. Phys. 141, 1982, 216 } 
\lref\klebtsey{ I. Klebanov, A. Tseytlin, 
``Entropy of Near-Extremal Black p-branes,'' 
hep-th/9604089, Nucl.Phys. B475 (1996) 164-178 }
\lref\bal{  A.P. Balachandran ``Quantum Spacetimes in the Year 1,''
hepth/0203259 } 
\lref\wal{ N.L. Wallach, ``Harmonic analysis on homogeneous spaces,'' 
M. Drekker Inc. NY 1973 } 
\lref\kramer{ M. Kramer, `` Some remarks suggesting an interesting 
 theory of harmonic functions on $SU(2n+1)/Sp(n)$ and 
$SO(2n+1)/U(n)$,'' Arch. Math. 33 ( 1979/80), 76-79. } 
\lref\fulhar{  W. Fulton and G. Harris,
``Representation theory,'' 
Springer Verlag 1991.}
\lref\bfss{ T. Banks, W. Fischler, S. Shenker, L. Susskind, 
``M-Theory as a Matrix model : A conjecture,'' 
  hep-th/9610043,  Phys.Rev.D55:5112-5128,1997 }
\lref\ramwal{ S. Ramgoolam, D. Waldram, 
``Zero branes on a compact orbifold,''  hep-th/9805191, 
 JHEP 9807:009,1998 }
\lref\grlayi{ 
Brian R. Greene, C.I. Lazaroiu, Piljin Yi
`` D Particles on $T^4 / Z(N)$ Orbifolds and their resolutions,''
hep-th/9807040,  Nucl.Phys.B539:135-165,1999 }
\lref\kazsuz{Y. Kazama, H. Suzuki,
 ``New $N=2$ superconformal field theories and superstring compactification''
 Nucl.Phys.B321:232,1989 }
\lref\hsy{ Yi-Xin Chen, Bo-Yu Hou, Bo-Yuan Hou
``Non-commutative geometry of 4-dimensional quantum Hall droplet,'' 
 hep-th/0203095 }


\def\Ga{ \Gamma } 
 
\def\cRn{ {\cal{R}}_n } 
\def\cRnp{ { { { \cal{R}}^+_n} }  } 
\def\cRnm{ { {{\cal{R}}^-_n}}  }
\def\cRnpm{ { { { \cal{R}}^{ \pm }_n} }  } 
\def\cRp{ { { \cal{R}}^+}  } 
\def\cRm{ { {\cal{R}}^-}  }
\def\Prp{ { {\cal{P}}_{\cal{R^+}} }  }

\def\Prnpm{ { \cal{P}}_{ {{\cal{R}}^{\pm}_n}}   }
\def\Prnp{ { \cal{P}}_{ {{\cal{R}}^+_n}}   }
\def\Prnm{ { \cal{P}}_{ {  {\cal{R}}^-_n} }      }
\def\Prn{  { \cal{P} }_{   {\cal{R}}_n    }   }

\newsec{ INTRODUCTION } 
 
 Fuzzy spheres provide many interesting solutions 
 to Matrix Brane actions \refs{ \kata, \clt, \myersdiel, \myrsnabint, \bmn}
 They  have also been suggested to play a role 
 in the context of a spacetime explanation of the stringy 
 exclusion principle \refs{ \malstrom, \jevram, \holi,\berkverl }.   

 A class of odd-dimensional fuzzy spheres was defined 
 in \guram\ and studied in more detail in \sphdiv. 
 The detailed $SO(m) $ decomposition of the 
 Matrix algebras related to the  fuzzy $S^{m}$ 
 was given in \sphdiv. For $m > 2 $  the Matrix algebras 
contain more representations than is necessary to describe 
functions on the sphere, and a projection is needed to 
 get the desired traceless symmetric representations.   
The geometry of general even dimensional fuzzy spheres 
 was studied in \hdim\ and it was found that the Matrix algebras
 related to the fuzzy sphere of $2k$ dimensions 
 approaches, in a limit of large matrices, 
 the algebra of functions on a
 higher dimensional space $ SO(2k+1)/U(k)$, the coset of
$SO(2k+1)$ by the right action of $U(k)$. This 
 lead to the statement, for $k \ge 2$, that fluctuations 
 around the fuzzy sphere solution can be described equivalently
 in two ways. On the one hand there is an abelian theory 
 on the higher dimensional coset. On the other hand there 
 is a non-abelian theory on the sphere. At finite $n$ this 
 non-abelian theory on the fuzzy sphere has to be formulated
 on a commutative but non-associative algebra \sphdiv.  
 The structure constants of the associative algebra 
 for the higher dimensional coset have been described more explicitly 
 in \hsy. 
 The abelian theory on the higher dimensional coset, in the case $k=2$ 
 has been further studied in \kimura. 
 A connection between the fuzzy four-sphere and 
 the 4D quantum Hall effect  \zhhu\ was found in \fab\ 
 and higher even fuzzy spheres were 
 used  to explore 6D and 8D generalizations of the quantum Hall
 effect. The extra degrees of freedom of the fuzzy $4$-sphere 
 have been discussed in the context of uncertainty relations in 
 \bal.

 In this paper we explore the extent to which 
 the picture of physics on even dimensional 
 fuzzy spheres developed in \hdim\ can be extended to 
 fuzzy odd spheres. We find that,  as in the even case, there is 
 an underlying higher dimensional geometry related to the 
 Matrix algebra. For the case of the sphere $S^{2k-1}$ 
 with symmetry $SO(2k)$, the relevant geometry 
 is ${SO(2k)\over U(k-1) \times U(1)}$, the coset of 
$SO(2k)$ by the right action of $U(k-1) \times U(1)$. 
 We will discuss in detail the cases $k=2$ and $k=3$ in this paper
 but it appears that arguments leading to these cosets 
 should generalize easily, especially the arguments 
 in section 3.3 and 8.1. 

 The precise role of the 
$ { SO(2k) \over U(k-1) \times U(1)} $ is, however, different 
from the role played by the $SO(2k+1)/U(k)$ in the 
 case of even dimensional spheres. 
 To explain this, we need to recall some details 
 about the construction of even and odd fuzzy spheres. 
  In all cases one starts with  matrices 
  $X_{i}$, where $i$ transforms in the vector of 
 $SO(m)$, and 
 \eqn\defeq{ \sum_{i=1}^{m} X_i^2 = 1 } 
 The $X_i$ are in $ End ( \cRn )$, i.e Matrices which are 
 transformations of the vector space $\cRn$, which is a representation 
of $SO(m)$. The $X_i$ in \defeq\ are related by a rescaling to 
 the $X_i$ used in the bulk of this paper.  
 $\cRn$ is a vector space whose dimension $N$ depends 
 on an integer $n$. The precise dependence of $N$ on $n$ 
  can be found in \sphdiv. The crucial difference 
 between even and odd spheres is that, in the case of 
 even spheres $ \cRn$ is an irreducible representation 
 of $SO(2k+1)$, whereas  in the case of odd spheres 
 $\cRn$ is a direct sum of irreducible representations 
 $ \cRnp $ and $ \cRnm$ of $SO(2k)$, that is, $ \cRn = \cRnp \oplus \cRnm $. 
 The matrices $X_i$ are maps from $ \cRnp $ to $\cRnm$ 
 and vice versa from $\cRnm$ to $\cRnp$.

 For even fuzzy spheres,  $  End ~ ( \cRn )$ becomes 
 commutative in the large $n$ limit and approaches the algebra of functions 
 on a classical space $ SO(2k+1) /U(k)$. 
 In the odd sphere case, both   $  End ( \cRnp )$
 and $  End ( \cRnm )$ become commutative in the large $n$ limit 
 and approach the algebra of functions
 on    $ {SO(2k) \over U(k-1) \times U(1) }   $. However 
 the full Matrix algebra $ End ( \cRn )$ remains 
 non-commutative. In the case $k=2$,  
 the  relevant higher dimensional space 
${SO(4) \over  U(1)  \times U(1) }$ has a simpler 
description as $S^2 \times S^2$ when 
 we use the isomorphism $ SO(4) \equiv SU(2) \times SU(2)$. 
 Section 3 gives two ways  to prove that $End( \cRnpm )$ 
 approaches the algebra of functions on $S^2 \times S^2$ in the large
 $N$ limit. 
 One uses the techniques of Kaluza-Klein reduction \sastr\ 
 or equivalently harmonic analysis on homogeneous spaces, as described
 for example in \wal.   
 Another uses the analysis of the stabilizer group 
 of a particular solution to the algebraic relations 
 satisfied by the generators of the Matrix algebra. 
 Section 8 proves, by similar methods, 
 that the higher dimensional geometry 
 related to the fuzzy five-sphere is ${SO(6) \over  U(2) \times U(1) }$.

 In sections 4-7 we explore the fuzzy 3-sphere 
 in more detail. Section 4 gives the 
 relations between the generators of $ End ( \cRn )$. 
 Section 5 gives the large $n$ limit of these 
 relations, showing, in particular, that 
 while $ End( \cRnpm )$ become commutative, 
 $ End ( \cRn ) $ remains non-commutative. 
 Section 6 uses the relations  derived in sections 4 and 5 to 
 construct some Matrix actions which are solved by the 
 fuzzy three-sphere Matrices. Section 7 
 discusses the physics of fluctuations around these 
 solutions. 
 
 In section 9.1 we discuss the geometry of these cosets 
 ${SO(2k) \over U(k-1) \times U(1)}$ further. In particular
 with describe some bundle structures they admit. 
 For the even sphere case, the bundle structure 
 of $SO(2k+1)/U(k)$ was used to map abelian field 
 theory on the fuzzy coset to non-abelian field theory 
 on the fuzzy sphere base \hdim. In the odd sphere case, the 
 relevant higher dimensional geometry 
 is not in general a bundle over the sphere, so the same strategy 
 as \hdim\ cannot be used to obtain a field theory 
 on the sphere. However some bundle structures 
 on the coset do exist and we describe them. An interesting 
 feature is that base and fibre are both hermitian symmetric spaces, 
 which come up in $N=2$ sigma models for example \kazsuz.  
 In section 9.3 we give a qualitative comparison  
 of some basic counting of degrees of freedom 
 which explains why it appears dificult to obtain 
 an ordinary  non-abelian theory on the fuzzy odd sphere. 
 We observe some intriguing similarities to discussions 
 of the entropy of M5-branes. 
 Section 9 also describes some avenues for future 
 research.

\newsec{ Review of fuzzy odd spheres } 

To get a fuzzy three sphere \guram\  we 
look for matrices which satisfy the equation
$\sum_{i} X_i^2 = c $, where $c$ is a constant. 
The matrices are constructed by starting 
with $V = V_+ \oplus V_-$. The vector 
space $V_+$ is the two-dimensional spinor 
representation of $SO(4)$ of positive chirality.  
The space $V_-$ is the two-dimensional representation
of negative chirality. The projector  $P_+$ acting 
on $V$ projects onto $V_+$, and $P_-$ projects 
onto $ V_-$. In terms of the $SO(4) = SU(2) \times SU(2) $ isomorphism, 
these have spins $ ( 2j_L, 2j_R ) = (1,0) $ 
and $  ( 2j_L, 2j_R ) = (0,1) $ respectively. 
For every odd  integer $n$, the 
space $ Sym ( V^{\otimes n } ) $ ( which is the symmetrized
$n$-fold tensor product space )  contains a
subspace where there are $ { (n+1)\over 2 }$ factors
of positive chirality and ${ (n-1) \over 2 } $ factors
of negative chirality. This subspace is an irreducible 
representation of $SO(4)$ labelled by  
$ ( 2j_L, 2j_R ) = ( { (n+1) \over 2 }  , { (n-1)\over 2 } ) $, 
which we will call $ \cRnp$. 
The projection operator in $ End ( Sym ( V^{\otimes n } ) ) $ 
which projects onto this irrep is called 
$\Prnp $. In \guram\ it was called $\Prp$, but here 
we are using notation which makes the $n$ dependence explicit. 
$ Sym ( V^{\otimes n } ) $ also contains a subspace 
where there are $ { (n+1)\over 2 }$ factors
of negative chirality and ${ (n-1) \over 2 } $ factors
of positive chirality. This subspace is an irreducible 
representation of $SO(4)$ labelled by  
$ ( 2j_L, 2j_R ) = ( { (n-1) \over 2 }  , { (n+1)\over 2 } ) $, 
which we will call $ \cRnm$.
The projector for this subspace 
is called  $\Prnm $. The space $ \cRn$ is defined to be a  
direct sum 
\eqn\defcrn{ 
 \cRn = \cRnp \oplus \cRnm } 
The projector for this space is a sum 
\eqn\defprn{ 
\Prn = \Prnp + \Prnm }
The Matrices $X_i $ are  linear transformations of 
$ \cRn $, i.e they are in $ End ( \cRn ) $. 
\eqn\fuzthree{  X_{i} = \Prn  ~ \sum_{r} \rho_r ( \Gamma_{i} ) ~  \Prn } 
More precisely, $X_i$ map $ \cRnp$ to $\cRnm$ 
and $\cRnm$ to $\cRnp$. This can be expressed 
by saying it is a sum of matrices in $Hom( \cRnp, \cRnm ) $
and in $Hom( \cRnp, \cRnm ) $  
\eqn\xassum{ 
X_i = \Prnp X_i \Prnm +  \Prnm X_i \Prnp } 
In other words, if we arrange the vectors of 
$\cRp$ along the upper rows of a column vector 
and those of $ \cRm$ along the lower rows, 
the matrices $X_i$ are non-zero in the off-diagonal blocks.

Let us prove that $ \sum_{i} X_{i}^2 $ commutes with 
generators of $SO(4)$. 
\eqn\comm{\eqalign{  
&[  \sum_{i} X_{i}^2 , ~~\Prn~\sum_{t} \rho_t ( \Gamma_k \Gamma_l )~  \Prn
] \cr 
&= \sum_{i} [ X_{i} , ~~\Prn~ \sum_{t} \rho_t ( \Gamma_k \Gamma_l ) ~ \Prn]
~ X_i \cr 
& + X_{i} [ X_{i} , ~~  \Prn~  \sum_{t} \rho_t ( \Gamma_k \Gamma_l ) ~
\Prn]
\cr 
& = \sum_{i} \Prn~ \sum_{r} [ \rho_r ( \Gamma_i ), ~~ \sum_{t}  
\rho_t ( \Gamma_k \Gamma_l ) ] ~ \Prn\sum_{s} \rho_{s}  ( \Gamma_{i} ) 
~ \Prn \cr 
& + \Prn~ \sum_{r} \rho_{r} ( \Gamma_i ) ~ \Prn~ [  \sum_s \rho_s
( \Gamma_i),  ~~ \sum_{t} \rho_{t} ( \Gamma_k \Gamma_l )] ~ \Prn \cr 
&= \Prn~ \sum_r \rho_r ( [\Gamma_i, ~~ \Gamma_k \Gamma_l ] ) ~ \Prn\sum_s
\rho_s( \Gamma_i ) ~ \Prn\cr 
& + \Prn~ \sum_r \rho_r( \Gamma_i ) ~ \Prn\sum_s
\rho_s ( [ \Gamma_i, ~~ \Gamma_k \Gamma_l ] ) ~  \Prn\cr 
&= \Prn~ \sum_r \rho_r( \Gamma_l \delta_{ik} - \Gamma_k \delta_{il} )
~ \Prn
  \sum_s \rho_s ( \Gamma_i ) \Prn\cr 
& + \Prn~ \sum_{r} \rho_r (\Gamma_i  ) ~ 
 \Prn\sum_s ( \rho_s( \Gamma_l ) \delta_{ik} - \rho_s( \Gamma_k )
\delta_{il} ) ~ \Prn\cr 
&= 0 \cr }}

Since $X_i^2$ commutes, we know it is a constant in each irrep. 
\eqn\consc{ X_i^2 = a_+ \Prnp + a_- \Prnm } 
Now consider the action on $X_i^2$ on $R_+$. 
We find  
\eqn\xsqpp{ X_{i}^2 \Prnp = \Prnp \sum_{r} \rho_r( \Gamma_i P_- )
 \Prnm \sum_{s} \rho_{s} ( \Gamma_i P_+ ) \Prnp  = a_{+} \Prnp}
and likewise : 
\eqn\xsqpm{ X_{i}^2 \Prnp = \Prnm \sum_{r} \rho_r( \Gamma_i P_+ )
\Prnp \sum_{s} \rho_{s} ( \Gamma_i P_- ) \Prnm  = a_{-} \Prnm }
Now consider the operation of exchanging $P_+$ with $P_-$, 
which leaves $X_i^2$ as defined in \fuzthree\ invariant. 
Now peform this operation on the equation \xsqpp\ to find 
\eqn\opxp{  X_{i}^2 \Prnm = \Prnm \sum_{r} \rho_r( \Gamma_i P_+ )
\Prnp \sum_{s} \rho_{s} ( \Gamma_i P_- ) \Prnm  = a_+ \Prnm   }
Comparing with \xsqpm\ we find that $a_+ = a_-$. 
Thus $X_i^2$ is proportional to the identity.

The radius is easily calculated as follows : 
\eqn\radform{ X_i^2 \Prnp = \sum_{r} \rho_r ( \Gamma_i \Gamma_i P_+ ) 
 + \sum_{r \ne s }  \rho_r ( \Gamma_i P_- ) \rho_s ( \Ga_i  P_+ ) } 
In the second term, the  index $s$ can take ${ n+1 \over 2 }$ vlues for a 
 non-zero answer and the index $r$ can take $ { (n-1) \over 2 }$
 values. Also we have $ \sum_i ( \Ga_i \otimes \Ga_i)  P_- \otimes P_+ 
 = 2 ( P_+ \otimes P_- ) $.
The first term gives $ {(n+1) \over 2} \times 4 $. 
 Adding these two contributions we get, 
${ (n+1) \over 2 } ( 4 + (n-1) )   = { ( n+1) ( n+3) \over 2}$.

Note that 
\eqn\dist{\eqalign{&
 \Prn   ~ \sum_{i} \sum_{r} \rho_r ( \Gamma_{i} ) 
 \sum_{s} \rho_s ( \Gamma_{i} )   ~ \Prn \cr 
&  = \sum_{i} X_{i}^2 + 
  \Prn ~ \sum_{i} \sum_{r} \rho_r ( \Gamma_{i} )   
( {\cal{P}}_{{\cal{R}}_{n++}} +
   {\cal{P}}_{ {\cal{R}}_{n--}} ) 
 \sum_{s} \rho_s ( \Gamma_{i} )   ~ \Prn  \cr} }
where $ {\cal{P}}_{{\cal{R}}_{n++}} $
 projects onto $ ( 2j_{L} , 2j_{R} ) = ({  n+3 \over 2 },
{ n-3\over 2 }  )$, and  ${\cal{P}}_{ {\cal{R}}_{n--}}$ 
projects onto $ ( 2j_{L} , 2j_{R} ) = ({  n-3 \over 2 },
{ n+3\over 2 }  )$. 
We should be careful to distinguish the LHS of \dist\ from 
$X_i^2$.

The above discussion generalizes easily 
to fuzzy five spheres \sphdiv. 
A similar calculation gives 
\eqn\radff{ \sum_i X_i^2 = { ( n+ 1 )\over 2 } ( n+ 2k -1 ) }

\newsec{ Fuzzy three-sphere and $S^2 \times S^2$  } 

 The Matrix Algebras related to fuzzy $S^m$, for $m > 2 $ 
contain more representations
 than required to give the algebra of functions on 
  a sphere \sphdiv. In the even sphere case, the 
 higher dimensional manifolds related to the Matrix algbras 
 have been identified \hdim. We recall some relevant facts here. 
 In the case of the even fuzzy 
 sphere $ S^{2k} $ it was just $ SO(2k+1) /U(k) $. 
 Several different proofs of this result were given. 
 The first proof follows immediately  once we know \sphdiv\  
 that the Matrix algebra contains every representation of 
 $SO(2k+1)$ with unit multiplicity. We then used the result of
 \kramer\ that the algebra of functions on 
 $ SO(2k+1) /U(k) $ has precisely this 
 property. In section 3.1 we will follow this line of argument, 
 i.e we will compare the $SO(4) $ decomposition 
 of the functions on $S^2 \times S^2$ to the representations 
 found in $ End ( \cRnpm ) $ at large $n$ and find agreement. 
 In section 3.2 we will develop an analogous argument 
 for  $ Hom ( \cRnp, \cRnm ) $ or  $ Hom ( \cRnm, \cRnp ) $,
 showing that the space of sections of a bundle on 
 $S^2 \times S^2$ agrees with the $SO(4) $ content 
 of these off-diagonal matrices. A second line of argument to relate 
 fuzzy even spheres to the appropriate coset will be 
 briefly described in section 3.3 and the analog 
 will be developed for the fuzzy three-sphere.

\subsec{ Spectrum of  $End ( \cRnpm ) $ and Functions on $S^2 \times S^2$
} 

 Rotations $L_{12} $ and $L_{34} $ form a 
 $U(1) \times U(1) $ subgroup of $SO(4)$. 
 In a self dual representation associated 
 with an $SO(4)$ Young diagram of row lengths 
 $(r_1, r_2)$ the eigenvalues of 
these generators on the highest weight vector are  $ r_1, r_2$. 
 In an anti-self dual representation 
 associated with the same Young diagram 
 the eigenvalues on the highest weight vector are $ (r_1, -r_2)$.
 The  combination $ L_{12} + L_{34}$ generates a $U(1) $ subgroup 
 of the left $SU(2) $ in the $SU(2) \times SU(2) $ 
 description of $SO(4)$. The 
 $SU_L(2)$ weight is, therefore,  described by $2J_L = L_{12} + L_{34} $. 
 The combination $ L_{12} - L_{34}$ generates a $U(1) $ subgroup 
 of the right $SU(2) $. The $SU_R(2) $ highest weight is, therefore, 
described by $ 2J_R = L_{12} - L_{34} $. 
For self-dual reps, we have then $ 2J_L = r_1 + r_2 $ 
 and $ 2J_R = r_1 - r_2 $, where $r_2$ is positive. 
 For antiself-dual reps,  $ 2J_L = r_1 - r_2 $ and 
 $ 2J_R = r_1 +  r_2 $ for positive $r_2$.

 To get the spectrum of representations
 of $SU(2) \times SU(2) $ present among the harmonics 
 of  $  { ( SU(2) \times SU(2) ) \over  ( U(1) \times U(1) ) } $
 we consider the representation of  
 $ SU(2) \times SU(2) $ induced from the trivial 
 representation of $ U(1) \times U(1)$. 
 By Frobenius duality, the multiplicity of 
 an irrep. of $ SU(2) \times SU(2) $ in the 
 induced rep. is equal to the multiplicity 
 of the identity rep. in the restriction of that irrep 
 to $U(1) \times U(1) $. This allows us to recover 
 the well-known fact that the spectrum of 
 reps is $ (2j_L , 2j_R ) $ with $ 2j_L$ and $2j_R$ 
 even. 

 Reps of highest weight $(r_1, r_2)$ with 
 $r_2$ positive ( i.e self-dual reps ) 
 and $r_1-r_2 $ even, is the same as the set of reps 
 with $2j_L = r_1 + r_2 = ( r_1 -r_2 ) + 2r_2 $ even, 
 and $ 2j_R = r_1 - r_2 $ even, with the restriction 
 that $ 2j_L \ge 2j_R $. 
 Reps of highest weight $(r_1, - r_2)$ with $r_2 $ 
 positive   ( i.e anti-self-dual reps ) and $r_1 - r_2$ 
 even is the same as the set described 
 by $ 2j_L$ even and $2j_R $ even with  $ 2j_R \ge 2j_L $. 
 These sets of reps. of $SO(4) $  are the ones 
 which appear in the $SO(4) $ decompositon  
 of $ End( \cRnp ) $  or $ End ( \cRnm ) $  at large $n$ \sphdiv. 
 We conclude that large $n$ $SO(4) $  decomposition of 
 $ End ( \cRnp ) $   or $ End ( \cRnp ) $  has the same representations 
 as the space of functions on $S^2 \times S^2$. 
 Since the structure constants of the algebra are determined by the
 Clebsch-Gordan coefficients, once we have established that 
 the Matrix algebra has the same repesentations 
 as $ Fun ( S^2 \times S^2 ) $ we know that it is the same
 algebra.

\subsec{ Spectrum of  $Hom(\cRnp, \cRnm )$ and sections on $S^2 \times
S^2$ } 

 In the large $n$ limit of  $Hom(\cRnp, \cRnm )$ ( or $Hom(\cRnm,
 \cRnp )$ ) 
 we have 
 self-dual reps associated with Young Diagrams 
 labelled by non-negative row lengths $(r_1,r_2)$ ( which have
 highest  weights 
 $(r_1,r_2)$ ), and obeying the condition $ r_1-r_2$ 
 odd. We also have antiself-dual reps.  associated with Young Diagrams 
 labelled by non-negative row lengths $(r_1,r_2)$ ( which have 
 highest weights $(r_1, - r_2)$ ), and obeying 
 $ r_1-r_2$ odd. These results are found in \sphdiv.

 This approaches a space of sections 
 on $ { SO(4) \over  ( U(1)  \times U(1)  ) } $. 
 Spaces of sections on this coset 
 are given by induced representations. 
 By the the Frobenius reciprocity theorem, 
 the multiplicity  of an irrep of $R$ of  $G$ in 
 the rep. induced from $H$ using the irrep. 
 $r$ of $H$ is equal to the multiplicity of 
 $r$ in the restriction of $R$ to $H$. 

 The desired space of sections 
 is the one obtained by inducing the irrep 
 of $ U(1)  \times U(1) $ which corresponds to 
 $ ( 1, 0 )$. 
 ( We could also work with $ (-1,0) $ and get the 
 same spectrum as $ Hom ( \cRnm, \cRnp )$ or  $ Hom ( \cRnp, \cRnm )$). 
The first $U(1) $ is generated by $L_{12} $
 and the second by $ L_{34}$. 
 Writing the product of two $U(1)$ in terms of generators 
 of the two  $SU(2) $ 
 factors in the $ SO(4) \equiv SU(2) \times SU(2) $ isomorphism, 
 the charges are $ (2j_L , 2j_R ) = (1,1)$

 Consider all reps of $SU(2) \times SU(2) $
 which restrict to  the $ (2j_L , 2j_R ) = (1,1)$
 of the $U(1) \times U(1) $ subgroup. 
  They are all the reps with $ ( 2j_L , 2j_R ) $ 
 odd and each contains the $(1,1)$ 
 with unit multiplicity and therefore 
 appears in the space of sections with unit multiplicity. 
 This is exactly the description of the spectrum 
 of $SO(4)$ reps in $ Hom ( \cRnp, \cRnm )$ from 
 \sphdiv\ descirbed above.

\subsec{ $S^2 \times S^2$ by analysis of Stabilizer group } 
 Another  strategy in to identify the higher domensional 
 coset in \hdim\ was to think 
 about the system of equations obeyed by $X_{\mu}$ and $X_{\mu \nu}$. 
 In particular we needed to find just one solution to the 
 system of equations, and prove that the subgroup of $SO(5)$ which
 kept 
 that solution invariant was $U(2)$. To find a solution, we took the
 expectation 
 value of the matrices $X_{\mu} $ and $X_{\mu \nu }$ in one 
 particular state, the $n$-fold tensor product of the $v_0$
 ( see appendix for notation ).   

 We will do something similar for the fuzzy 3-sphere. 
 Identify a set of generators for the complete Matrix 
 algebra, and take expectation values in the large $n$ limit, 
 and analyze the stabilizer group if the expecttaion values.

 A  complete set of generators is : 
\eqn\gens{\eqalign{  
&  X_{i}^+ = \Prnm  ~ \sum_r \rho_r ( \Gamma_iP_+) ~  \Prnp    \cr 
&  X_{i}^{-} = \Prnp ~  \sum_r \rho_r ( \Gamma_i P_- ) ~ \Prnm   \cr 
&  X_{ij}^{+} =  \Prnp ~ \sum_r \rho_r (  
{ 1 \over 2 } [ \Gamma_i,  \Gamma_j ]  P_+ )~   \Prnp
\cr 
&  Y_{ij}^{+}  = \Prnp ~ \sum_r \rho_r ( { 1 \over 2 } 
[ \Gamma_i,  \Gamma_j]  P_- )~  \Prnp
\cr 
&  X_{ij}^{-} =  \Prnm  ~ \sum_r \rho_r ( { 1 \over 2 }
[  \Gamma_i , \Gamma_j] P_- ) ~ 
\Prnm  \cr 
& Y_{ij}^{-}  = \Prnm ~  \sum_r \rho_r ( { 1 \over 2 }
 [ \Gamma_i,  \Gamma_j ] P_+ )~    \Prnm 
\cr }}

The coordinates of the sphere $X_i$ are related 
to the above by 
$X_{i} = X_{i}^{+} + X_{i}^{-} $. 
It is also useful to define 
\eqn\mdefs{\eqalign{
&  X_{ij} = X_{ij}^{+} + X_{ij}^{-} \cr  
&  Y_{ij} =  Y_{ij}^{+} + Y_{ij}^{-} \cr 
& Y_i = X_i^+ - X_i^- \cr 
& \tilde X_{ij} =   X_{ij}^{+} -  X_{ij}^{-} \cr 
& \tilde Y_{ij} =  Y_{ij}^{+} - Y_{ij}^{-}
}}

Take a state $ |s> $  in $ \cRnp $ 
containing ${ ( n+1) \over 2 }$ copies of 
$ v_0$ and $  { ( n-1) \over 2 }$ copies 
of $ a_1^{\dagger} v_0 $. We find that 
\eqn\expvals{\eqalign{  
& < s | { i~ X_{12}^{+} \over n }  |s >   =   -{ 1   \over 2 }  \cr 
& < s | { i ~ Y_{12}^{+} \over n}   |s>   =   - { 1   \over 2 }   \cr 
& < s |  { i X_{34}^{+}\over n }  |s >  =   - {1 \over 2 }     \cr 
&  < s| { -i Y_{34}^{+} \over n }  |s>    =  { 1 \over 2 }  \cr  }}
Other expectations values are zero. 
The algebraic equations obeyed 
by $ X_{ij}^{+}$ and $ Y_{ij}^{+}$ 
have a solution with the above numerical values.

These values are stabilized by 
 the rotations $ L_{12}$ and $L_{34} $ 
 which generate a $U(1) \times U(1) $ subgroup 
 of $SO(4) $. The $SO(4) $ acts transitively on the 
 solutions, so the set of solutions is $ {SO(4) \over U(1) \times
 U(1)}$.

\newsec{ Summary of relations between generators }

 There are the sphere-like relations  which generalize 
 the defining equation of the 3-sphere.  
\eqn\sphrl{\eqalign{ 
& \sum_i  X_i X_i  = { ( n +1 ) ( n+3) \over 2 }  \cr 
& \sum_{i \ne j} X_{ij}^+ X_{ij}^+  =  - ( n +1 ) ( n+5 )  \cr 
&  \sum_{i\ne j} X_{ij}^- X_{ij}^- =  - ( n +1 ) ( n+5 )   \cr 
& \sum_{i \ne j } Y_{ij}^+ Y_{ij}^+  = - ( n -1 ) (n+ 3)   \cr
 & \sum_{i \ne j} Y_{ij}^- Y_{ij}^- = -  ( n - 1 ) ( n + 3 )\cr }}

There are self duality or antiself-duality relations 
\eqn\dltrls{\eqalign{ 
& X_{ij}^+ = { 1 \over 2 }  \epsilon_{ijkl } X^+_{kl} \cr 
& X_{ij}^- = - { 1 \over 2 }  \epsilon_{ijkl } X^-_{kl} \cr 
& Y_{ij}^+  = - { 1 \over 2 }  \epsilon_{ijkl } Y^-_{kl} \cr 
& Y_{ij}^- =   { 1 \over 2 } \epsilon_{ijkl } Y^-_{kl} \cr }}

Incidentally,  the (anti)self-duality equations show that 
 there are only 3 independent components 
 for $X_{ij}^+$ and 3 for $Y_{ij}^{+} $. 
 We are using conventions $ \Gamma_5 = - \Gamma_1 \Gamma_2 \Gamma_3
 \Gamma_4$ with $ \Gamma_5 v_0 = v_0 $.   
Together with the equations in \sphrl\ 
 this shows explicitly the origin of the 
 $S^2 \times S^2 $ in the upper block. 
 Similarly there is the $S^2 \times S^2$ 
 in the lower block defined by the 
  $X_{ij}^-$ and  $Y_{ij}^{-} $ variables.

Consider multiplying $X_i^{\pm}$ with $X_j^{\pm} $ 
\eqn\relsxx{\eqalign{
& ( X_{i}^{+} X_{j}^- -  X_{j}^{+} X_{i}^- ) 
  + { \epsilon_{ijkl} \over 2 } ( X_{k}^{+} X_{l}^- -  X_{l}^{+} X_{k}^- )
 = - {(n-1)\over 2 } \epsilon_{ijkl} X_{kl}^{-} + 2 ( X_{ij}^{-} + 
{ \epsilon_{ijkl} \over 2 } X_{kl}^{-} )   \cr 
& ( X_{i}^{+} X_{j}^- -  X_{j}^{+} X_{i}^- )
- { \epsilon_{ijkl} \over 2 } ( X_{k}^{+} X_{l}^- -  X_{l}^{+} X_{k}^- )
= - { n+1 \over 2 } \epsilon_{ijkl} Y^-_{kl} + 2 ( X_{ij}^- - 
{ \epsilon_{ijkl} \over 2 } X_{kl}^{-} ) \cr }}

By converting all $+$ to $-$ and at the same time 
converting $\epsilon_{ijkl} $ to $ - \epsilon_{ijkl}$, 
 we can write the folllowing equation : 
\eqn\rsxx{\eqalign{
& ( X_{i}^{-} X_{j}^+ -  X_{j}^{-} X_{i}^+ ) 
  - { \epsilon_{ijkl} \over 2 } ( X_{k}^{-} X_{l}^+ -  X_{l}^{-} X_{k}^+ )
 =  {(n-1)\over 2 } \epsilon_{ijkl} X_{kl}^{+} + 2 ( X_{ij}^{+} - 
{ \epsilon_{ijkl} \over 2 } X_{kl}^{+} )   \cr 
& ( X_{i}^{-} X_{j}^+ -  X_{j}^{-} X_{i}^+ )
+ { \epsilon_{ijkl} \over 2 } ( X_{k}^{-} X_{l}^+ -  X_{l}^{-} X_{k}^+ )
= { n+1 \over 2 } \epsilon_{ijkl} Y^+_{kl} + 2 ( X_{ij}^+ + 
{ \epsilon_{ijkl} \over 2 } X_{kl}^{+} ) \cr }}

From these we can write down 
\eqn\txcom{\eqalign{
&  ( X_{i}^{-} X_{j}^+ -  X_{j}^{-} X_{i}^+ ) = { n\over 4 }
\epsilon_{ijkl} 
( X^{+}_{kl}+  Y^+_{kl} ) + { 1 \over 4 } \epsilon_{ijkl} (- X^+_{kl} +
Y^+_{kl} ) + 2 X_{ij}^+ \cr 
&  ( X_{i}^{+} X_{j}^- -  X_{j}^{+} X_{i}^- ) =  - { n\over 4 }
\epsilon_{ijkl} 
( X^{-}_{kl} + Y^-_{kl} ) + { 1 \over 4 } \epsilon_{ijkl} ( X^-_{kl} -
Y^-_{kl} ) + 2 X_{ij}^- \cr }}

By adding the two equations above we get  : 
\eqn\xcom{\eqalign{
  [ X_i , X_j ] &= - {n\over 4} \epsilon_{ijkl} \bigl[ ( X^-_{kl} -
X^{+}_{kl} ) + ( Y^{-}_{kl} - Y^+_{kl} ) \bigr] \cr 
&  -{ 1\over 4 } \epsilon_{ijkl} \bigl[ ( X_{kl}^+ - X_{kl}^- ) -
( Y_{kl}^+ - Y_{kl}^- ) \bigr] + 2 X_{ij} \cr 
&  = {   n\over 4} \epsilon_{ijkl} (  \tilde X_{kl} +   \tilde
Y_{kl} )  -  {1\over 4 } \epsilon_{ijkl} 
( \tilde X_{kl} -  \tilde Y_{kl} ) + 2 X_{ij} \cr }}

By taking the difference of the two eqs. in \txcom\ 
we find 
\eqn\acom{ 
\{ X^{+}_i, X^-_j \} - \{ X^{+}_j, X^{-}_{i} \} 
 = - { n \over 4 } \epsilon_{ijkl} ( X_{kl} + Y_{kl} )
   + { 1 \over 4 } \epsilon_{ijkl} (  X_{kl}-  Y_{kl} ) 
- 2 ( \tilde X_{ij} ) } 
This can be expressed as a commutator of $X_{i} $ and $Y_i$. 
\eqn\acomi{ \{   X_i, Y_j \} = 
{ n \over 4 } \epsilon_{ijkl} ( X_{kl} + Y_{kl} )
     - { 1 \over 4 } \epsilon_{ijkl} ( X_{kl}-  Y_{kl} ) 
 + 2 ( \tilde X_{ij} ) }

We of course have relations like 
\eqn\nilpot{\eqalign{
& X_i^+ X_j^+ = 0 \cr 
& X_i^- X_j^- = 0 \cr }}
which follow from 
\eqn\ppzr{ 
\Prnp \Prnm = \Prnm \Prnp = 0 }

Now we consider multiplying $X_{ij}^{\pm}  $ 
and $Y_{ij}^{\pm}  $ 
with $X_{k}^{\pm} $ 

\eqn\xiijxk{\eqalign{ 
& X_{kl}^+ X_i^- - X_i^-Y_{kl}^- = \delta_{il} X_k^- -
\delta_{ik}X_l^- \cr 
& X_{kl}^{-} X_{i}^+ - X_i^+ Y_{kl}^{+} = \delta_{il} X_k^+ -
\delta_{ik} X_l^+ \cr 
& Y_{kl}^- X_i^+ - X_i^+ X_{kl}^+ = \delta_{il} X_{k}^- - \delta_{ik}
X_{l}^- \cr 
& Y_{kl}^+ X_i^- - X_i^- X_{kl}^- = \delta_{il} X_k^+ -
\delta_{ik}X_l^{+} \cr }}

Defining $C_{kli}^j \equiv \delta_{il} \delta_{kj} - \delta_{ik}
\delta_{jl}$ we can rewrite the above as : 
\eqn\xiijxki{\eqalign{ 
& X_{kl}^+ X_i^- - X_i^-Y_{kl}^- =  C_{kli}^j X_{j}^-  \cr 
& X_{kl}^{-} X_{i}^+ - X_i^+ Y_{kl}^{+} =  C_{kli}^j X_{j}^+ \cr 
& Y_{kl}^- X_i^+ - X_i^+ X_{kl}^+ =      C_{kli}^j X_{j}^+    \cr 
& Y_{kl}^+ X_i^- - X_i^- X_{kl}^- =  C_{kli}^j X_{j}^- \cr }}

Combining these with the 
 facts : 
\eqn\obvfcts{\eqalign{
&  X_{kl}^+ X_i^+ = X_i^+ X_{kl}^- =  0 \cr 
&  X_{kl}^- X_i^- =  X_i^+ X_{kl}^- = 0 \cr }}
which follow from \ppzr\ 
these imply that $X_{kl}^{\pm }  + Y_{kl}^{\pm}$ acts 
by commutators as generators of $SO(4)$ rotations.   
\eqn\comacts{\eqalign{ 
& [ X_{kl}^+ + Y_{kl}^- , X_i^- ] =   C_{kli}^j X_{j}^- \cr 
& [ X_{kl}^- + Y_{kl}^+ , X_i^+ ] =  C_{kli}^j X_{j}^+ \cr 
& [ X_{kl}^+ + Y_{kl}^-, X_i^+ ] =  C_{kli}^j X_{j}^+ \cr 
&  [ X_{kl}^- + Y_{kl}^+, X_i^- ] =   C_{kli}^j X_{j}^- \cr }}
The differences    $X_{kl}^{\pm }  - Y_{kl}^{\pm}$ 
act by anticommutators as a rotation combined with 
change of $X_{i}^+ $ to $X_i^-$ : 
\eqn\acmacts{\eqalign{ 
&  \{  X_{kl}^+ - Y_{kl}^- , X_i^- \}  =   C_{kli}^j X_{j}^- \cr 
&   \{ X_{kl}^- - Y_{kl}^+ ,  X_i^- \} =  - C_{kli}^j X_{j}^- \cr 
&   \{  X_{kl}^- - Y_{kl}^+ , X_{i}^+ \}  =  C_{kli}^j X_{j}^+ \cr 
& \{  X_{kl}^+ - Y_{kl}^- , X_{i}^+ \} =  - C_{kli}^j X_{j}^+ \cr }}
Recalling from \mdefs\  that $X_i = X_i^+ + X_i^-$ 
 and  $Y_i = X_i^+ - X_i^-$  allows 
us to write some interesting anti-commutator actions : 
\eqn\acts{\eqalign{ 
& \{ X_{kl}^+ - Y_{kl}^- , X_i \} = -  C_{kli}^j Y_j \cr 
& \{ X_{kl}^- - Y_{kl}^+ , X_i \} =   C_{kli}^j Y_j \cr
& \{ X_{kl}^+ - Y_{kl}^- , Y_i \} = -  C_{kli}^j X_j \cr 
& \{ X_{kl}^- - Y_{kl}^+ , Y_i \} =  C_{kli}^j X_j \cr 
 }}

It follows that : 
\eqn\pvnopm{\eqalign{& 
 [X_{kl} + Y_{kl} , X_i ] = 2  C_{kli}^j X_{j} \cr
& [X_{kl} + Y_{kl} , Y_i ] = 2  C_{kli}^j Y_{j} \cr 
 & \{ X_{kl} - Y_{kl} , X_i \} = 0 \cr 
& \{ X_{kl} - Y_{kl} , Y_i \} = 0 \cr 
}}
Recalling from \mdefs\ that  $\tilde X_{kl} = X_{kl}^+ - X_{kl}^- $ 
and $ \tilde Y_{kl} = Y_{kl}^+ - Y_{kl}^- $  we have 
\eqn\tildgen{\eqalign{ 
&  \{ \tilde X_{kl}  + \tilde Y_{kl},  X_i \}  = -2   C_{kli}^j Y_j 
\cr 
&  \{ \tilde X_{kl}  + \tilde Y_{kl},  Y_i \}  = - 2   C_{kli}^j X_j
\cr 
&  [ \tilde X_{kl}  - \tilde Y_{kl} , X_i ] = 0 \cr 
&  [ \tilde X_{kl}  - \tilde Y_{kl} , Y_i ] = 0 \cr
}}

So we have interesting anticommutator actions which 
are $SO(4) $ rotations combined with an $X-Y$ flip.

Finally we need to consider multiplication 
$ X_{ij}X_{kl} $ and $X_{ij}Y_{kl} $ 
Some of these are familiar from the fuzzy $S^4$ case : 
\eqn\xxijkl{\eqalign{& 
 [X_{ij} , X_{kl } ] = \delta_{jk} X_{il} + \delta_{il} X_{jk} 
 -  \delta_{jl} X_{ik}  -  \delta_{il} X_{jk } \cr 
& [Y_{ij} , Y_{kl } ] = \delta_{jk} Y_{il} + \delta_{il} Y_{jk} 
 -  \delta_{jl} Y_{ik}  -  \delta_{il} Y_{jk } \cr }}
 
Further relations are 

\eqn\gs{\eqalign{ 
& [ X_{ij}, Y_{kl} ] = 0 \cr 
&  X_{(ij}^+  Y_{kl)}^+ = - \delta_{jk} ( X_{i}^{-} X_{l}^+  + 
 X_{l}^{-} X_{i}^+ ) -  
   \delta_{il} ( X^-_j X^+_k + X^-_k X^+_j ) \cr 
&       + \delta_{jl} ( X^-_i X^+_k + X^-_k X^+_i  ) 
 + \delta_{ik} ( X^-_j X^+_l + X^-_l X^+_j  ) \cr 
 & + { ( n+ 1 ) (n+3) \over 2 } 
 ~ ( -\delta_{ik} \delta_{jl} + \delta_{il} \delta_{jk} ) \Prnp \cr 
& X^{+}_{ij} Y^{+}_{il} +  X^{+}_{il} Y^{+}_{ij}
= 2 ( X^-_jX^+_l  + X^-_lX^+_j) - { (n+1) (n+3) \over 2 } \delta_{jl} \Prnp \cr 
}}  

By taking $ +$ to $-$ we have 
\eqn\gsii{\eqalign{
& X_{(ij}^-  Y_{kl)}^- = - \delta_{jk} ( X_{i}^{+} X_{l}^-  + 
 X_{l}^{+} X_{i}^- ) -  
   \delta_{il} ( X^+_j X^-_k + X^+_k X^-_j ) \cr 
&       + \delta_{jl} ( X^+_i X^-_k + X^+_k X^-_i  ) 
 + \delta_{ik} ( X^+_j X^-_l + X^+_l X^-_j  ) \cr
& X^{-}_{ij} Y^{-}_{il} +  X^{-}_{il} Y^{-}_{ij}
= 2 ( X^+_jX^-_l +  X^+_lX^-_j )    -
 { (n+1) (n+3) \over 2 } \delta_{jl} \Prnm \cr 
}}  

This implies 
\eqn\gsi{\eqalign{ 
   X_{(ij}  Y_{kl)} & =  \tilde X_{(ij} , \tilde Y_{kl)}  \cr 
 & = - \delta_{jk}  \{ X_i, X_l  \}  -  \delta_{il}  \{ X_j, X_k  \} 
  + \delta_{jl}   \{ X_i, X_k  \}  + \delta_{ik} \{ X_j, X_l   \} \cr  
&   + { (n+1) (n+3) \over 2 }  ( -\delta_{ik} \delta_{jl} + \delta_{il} \delta_{jk} )
\cr }}
  
A special case of \gsi\ is 
\eqn\gsii{ 
   X_{(ij}  Y_{il)}  =  2 \{ X_j , X_l \} - 
{(n+1)(n+3) \over 2 } \delta_{jl} = -2     \{ Y_j , Y_l \} - 
{(n+1)(n+3) \over 2 } \delta_{jl} } 
It is worth 
pointing out that the equations of this section 
correct some equations in the appendix of \guram\ 
and also the remarks in section 7.3 of \sphdiv.

\newsec{ Algebra of fuzzy $3$-sphere at  large $N$ }

It is useful to define, as in the case of 
fuzzy even spheres, rescaled variables. 
A surprising feature will be that, 
after the standard rescalings, some non-commutativity 
wil remain at large $n$ in the fuzzy odd cases.

We can define variables 
\eqn\largen{\eqalign{
& A^{\pm}_{ij} = { i X_{ij}^{\pm} \over n}  \cr 
&  B^{\pm}_{ij} = {i  Y_{ij}^{\pm} \over n } \cr 
& C_{i}^{\pm} = { \sqrt{2} X_i^{\pm} \over n } \cr }}

 and the following linear combinations are useful.
\eqn\combs{\eqalign{ 
& A_{ij} = A_{ij}^{+}  + A_{ij}^{-} \cr 
& {\tilde A_{ij}}  =    A_{ij}^{+}  - A_{ij}^{-} \cr 
&  B_{ij} = B_{ij}^{+}  +  B_{ij}^{-} \cr 
& { \tilde B_{ij} } = B_{ij}^{+}  -  B_{ij}^{-} \cr 
& C_i =  ( C_i^+ + C_i^- )  \cr 
& \tilde C_i =  i  ( C_i^+ - C_i^- )  \cr }}

The sphere-like relations are of the form : 
\eqn\sphlgn{\eqalign{ 
& \sum_{i \ne  j } ( A_{ij}^{\pm } )^2  = \Prnpm     \cr 
& \sum_{i \ne j }  ( B_{ij}^{\pm} )^2   =  \Prnpm \cr 
& \sum_{i \ne j } ( A_{ij} )^2 = 1 \cr 
& \sum_{i \ne j } ( B_{ij} )^2 =1 \cr
& \sum_{ i }  C_{i}^2  = 1  \cr
& \sum_{i} \tilde C_{i}^2 = 1 \cr  }}

We have the following vanishing commutators and 
anticomutators : 
\eqn\vanrsl{\eqalign{ 
& [ A , A ]  = 0 \cr 
& [ B, B ] = 0 \cr 
& [ A, C ] = [ A, \tilde C ]   =0 \cr 
& [ \tilde A + \tilde B , C ] = [ \tilde A + \tilde B , \tilde C ] = 0
\cr 
& \{   \tilde A - \tilde B , C \}  = \{  \tilde A - \tilde B , \tilde
C \}  = 0   
\cr }}

There are  also some interesting commutators 
of $C, \tilde C$ which survive in the large N limit  
\eqn\survi{\eqalign{
&  ( C_{i}^{-} C_{j}^+ -  C_{j}^{-} C_{i}^+ ) = { -i\over 2 }
\epsilon_{ijkl} 
( A^{+}_{kl}+  B^+_{kl} ) \cr 
&  ( C_{i}^{+} C_{j}^- -  C_{j}^{+} C_{i}^- ) =   {  i  \over 2 }
\epsilon_{ijkl} ( A^{-}_{kl} + B^-_{kl} )   \cr }}

\eqn\surv{\eqalign{ 
 [ C_i , C_j ] & = { - i \over 2 } ~ \epsilon_{ijkl} ~( \tilde A_{kl} + \tilde
B_{kl} ) \cr 
  \{ C_i, \tilde C_j \} & =  { -1 \over 2 } ~ \epsilon_{ijkl}~ (   A_{kl} +
B_{kl} )  \cr 
 A_{(ij} ~ B_{kl)}  & = ~   - \{ \tilde A_{ij} , \tilde B_{kl} \} \cr 
& =  { \delta_{jk} \over 2 }  \{ C_i, C_l  \}  +  
{ \delta_{il} \over 2}  \{ C_j, C_k  \} 
  - { \delta_{jl} \over 2}    \{ C_i, C_k  \}  - { \delta_{ik}\over 2} 
 \{ C_j, C_l   \} \cr  
&  ~~~  + {1\over 3}
~~ ( \delta_{ik} \delta_{jl} - \delta_{il} \delta_{jk} ) \cr
& =  + { \delta_{jk} \over 2}  \{ \tilde C_i, \tilde C_l  \}  + 
{ \delta_{il} \over 2}  \{ \tilde C_j, \tilde C_k  \} 
   - { \delta_{jl}\over 2}    \{ \tilde C_i, \tilde C_k  \}  - 
 { \delta_{ik} \over 2 }  \{ \tilde C_j, \tilde C_l   \} \cr  
&   ~~~ + { 1\over 3}  ~~  ( \delta_{ik} \delta_{jl} - \delta_{il} \delta_{jk} ) \cr
  A_{( ij} ~ B_{il) }  & = - \{ C_j, C_l \} + {\delta_{jl} \over 2} \cr 
}}

So $A,B$ should be thought of as coordinates
and $C$ could perhaps be realized as derivatives. 
This is compatible with the observation based on 
group theory, that the diagonal blocks are 
describing functions on two copies of $S^2 \times S^2$, 
while the off-diagonal are some sections of a bundle. 
The fact that within each diagonal block $End( \cRnp)$ or 
$End( \cRnm)$ the 
algebra becomes commutative in the large $n$ 
limit, whereas there is non-commutativity 
in $ End ( \cRn )$ suggests that  some sort 
of $U(2)$ non-abelian structure is present. 
We will return to this in section 7.

\newsec{ Some Matrix Actions which admit the fuzzy three-sphere as
solution } 

It was observed in \kimura\ that the 
fuzzy two-sphere and four sphere 
are solutions to Matrix brane actions which 
have a mass term. Such matrix models in relation to fuzzy 
spheres were also discussed in \hosph. 
Mass terms also appear in 
the BMN matrix model and the 
fuzzy two-sphere  appears as a supersymmetric 
solution of the BMN matrix model \bmn. 
It is natural to ask if a simple Matrix Model 
with mass term can accommodate the fuzzy three sphere 
as a solution. In the case of the fuzzy even spheres 
the anlog of \xcom\ is much simpler, since it only 
contains the analog of the last term on the RHS of \xcom. 
Further, the commutator $ [X_{i} , X_{ij} ] $ is proportional 
to $ X_j$ in the even case,
 unlike the more complicated analog \pvnopm\ which 
involves both $X$ and $Y$ variables. These properties of 
the fuzzy even spheres are adequate to prove they are
solutions of a matrix model with $X_i$ in the vector 
of the $SO$ symmetry group and having a mass term. 
The form of the equations for the fuzzy three-sphere 
do not allow it to be a solution of such a Matrix model. 
It is a very interesting problem to try to find  {\it any } 
physical Matrix model where the fuzzy three-sphere  ( or the  fuzzy 
five sphere discussed later in this paper ) is a solution.
It would be even more fascinating to find these general 
fuzzy spheres as  supersymmetry-perserving solution 
of any physical Matrix model.  
A  physical Matrix model 
would be one that comes up as an action for branes 
( parallel or intersecting ) or as a dual to some 
M-theory background. In the light of the difficulty 
of constructing a transverse five-brane from  the BFSS 
matrix model, finding such a fuzzy five sphere as
a physical Matrix model solution ( or proving it is not possible ) 
would give important information about the relation between M-theory
and large $N$ matrix systems. 
We will not solve this problem in this paper but 
we will exhibit some toy Matrix models 
which do have the fuzzy three sphere as a solution, 
and we will observe some qualititative similarities 
between these toy Matrix models and some others that have appeared 
in recent literature.

Given the relations we wrote down in \xcom\ and 
\acom\ it is easy to obtain some Matrix Actions 
which admit as solutions the matrices of the 
fuzzy three-sphere. 
Consider a Matrix action with 
a symmetry group which contains $SO(4))$
and consider fields transforming in the vector of
$SO(4)$, labelled $\Phi_i$ as in the BFSS \bfss\ or IKKT \ikkt\vipul\  
matrix models. Suppose, unlike these models there is an  
additional Matrix variable transforming in the antisymmetric of 
$SO(4) $ which we will denote $P_{ij} $.  
Suppose the action is 
\eqn\actphp{ 
\int TR ~~ \bigl( ~ [ \Phi_i, \Phi_j ] -  P_{ij} ~ \bigr)^2 } 
The variation of this action with respect to $\Phi_i$ 
or $ P_{ij}$ will be proportional to 
$$ [ \Phi_i, \Phi_j ] - P_{ij}$$
From \xcom\ it follows  that if we set
\eqn\sol{\eqalign{  
& \Phi_{i } = X_{i} \cr 
& P_{ij} =   {   n\over 4} \epsilon_{ijkl} (  \tilde X_{kl} +   \tilde
Y_{kl} )  -  {1\over 4 } \epsilon_{ijkl} 
( \tilde X_{kl} -  \tilde Y_{kl} ) + 2 X_{ij}
 }}
where $X_i, X_{kl}, Y_{kl}, \tilde X_{kl}, \tilde Y_{kl} $ 
are defined in \gens\mdefs,  
we will have a solution to the action \actphp. 

Another way to construct an action which is solved 
by the matrices of section 3 is to 
start with variables $ \Phi_i$ and $\tilde \Phi_{i}$ 
and antisymmetric Matrix variables $ Q_{ij} $ 
which have an action 
\eqn\actpptq{ 
\int ~~ TR ~~  ( ~ \{  \Phi_i, \tilde \Phi_j \}  -  Q_{ij} ~  )^2 }
In this case if we use 
\eqn\asol{\eqalign{  
& \Phi_i = X_i \cr 
& \tilde \Phi_i = Y_i \cr 
& Q_{ij} =  { n \over 4 } \epsilon_{ijkl} ( X_{kl} + Y_{kl} )
     - { 1 \over 4 } \epsilon_{ijkl} ( X_{kl}-  Y_{kl} ) 
 + 2 ( \tilde X_{ij} ) \cr }}

It is interesting that in recent literature 
Matrix actions have been considered
which involve matrices transforming in the 
vector of the $SO$ group as well as other antisymmetric 
tensors \chaudhuri\bilal\smol\bl. It would be interesting 
to see if the fuzzy odd sphere solutions 
of  model actions of the form \actphp\ 
could also be found for actions involved in the above papers. 
Perhaps relating terms of the form in \actphp\ 
to terms in these actions could be useful. 
The role of $P_{ij}$ could also be played by 
composite fields such as $ \bar \psi \Gamma_{ij} \psi $.

\newsec{Fluctuating fields around solution } 

\subsec{ Physics on $S^2 \times S^2 $ } 
 
Useful information about the physics of a solution 
defined by the $X_i$ matrices is obtained 
by considering  fluctuations around such a solution. 
In the even sphere case the action for the fluctuations is a  
$U(1) $ theory on $SO(2k+1)/U(k) $ with 
$SO(2k+1) $ symmetry.  The existence of a hidden 
higher dimensional coset in the odd sphere allows 
a somewhat analogous result. 

Imagine we find a solution to Matrix theory 
which uses these matrices 
\eqn\mats{\eqalign{ 
&  \Phi_i = X_{i} \cr 
&  \Phi_a = 0 \cr }}
The $ \Phi_i$ transform as a vector of the $SO(4) $ symmetry group 
of the fuzzy three sphere, and we call them parallel scalars.
The $\Phi_a$ are invariant under that 
symmetry group and are called transverse scalars. 

Consider the fluctuations  
\eqn\flct{\eqalign{ 
& \Phi_i = X_i +  \phi_i^+ ( A^+, B^+ )   + \phi_i^{-} ( A^-, B^- ) 
           + \phi_{ij}^+ ( A^+, B^+ ) C_i^-  + \phi_{ij}^- ( A^-, B^-) C_j^+ 
\cr 
& \Phi_a = \phi_{a}^+ ( A^+, B^+ ) + \phi_{a}^- ( A^-, B^- )  
    + \phi_{ai}^+ ( A^+, B^+ ) C^-_i +   \phi_{ai}^- ( A^-, B^- ) C^+_i
\cr }}

Each $\phi$ is a field living on $ S^2 \times S^2$. 
Notice that the transverse scalars contain 
components which transform as vectors under the 
parallel $SO(4)$. The parallel scalars 
give rise to fields  $ \phi^{\pm}_{ij} $ which is a two-index tensor 
which can be reduced to symmetric traceless part, antisymmetric 
part and a scalar.
By expanding a Matrix action which is solved 
by the fuzzy three-sphere Matrices, around the solution, 
we would get a field theory, in analogy to the analogous 
results for fuzzy even spheres \iktw\hdim\kimura. 

We  outline some features 
of the field theory on $S^2 \times S^2$ that results
from these fluctuations. 
The kinetic term of $ \Phi_a  $ is 
\eqn\kint{ 
\int dt  ~TR~  ( \partial_t \Phi_a \partial_t  \Phi_a )  } 
Using the expansion for $\Phi_a$ in \flct, 
we will get, from the diagonal terms,   
\eqn\kinti{\eqalign{&  
 \int dt dA^+ dB^+  \bigl( \partial_t \phi_a^{+} ( A^+, B^+)  \bigr)^2 
+ \int dt dA^- dB^-   \bigl(  \partial_t  \phi_a^{-} ( A^-, B^- )  \bigr)^2 
\cr 
& 
= \int dt dA dB  \bigl( \partial_t \phi_a^{+} ( A, B ) \bigr)^2
  + \bigl( \partial_t \phi_a^{-} ( A, B ) \bigr)^2  \cr  }}    
$dA^+ dB^+ = dA^- dB^- $ is an $SO(4)$ invariant measure 
for $S^2 \times S^2$. These diagonal terms coming from the large 
$n$ limit of 
$End ( \cRnp ) $ and $ End ( \cRnm ) $.   
From squaring the off diagonal terms, 
$\phi_{ai}^+ ( A^+, B^+ ) C^-_i +   \phi_{ai}^- ( A^-, B^- ) C^+_i$
we obtain terms including  
\eqn\offkin{ 
\int dt ~ TR_{\cRnp}  ~  ( \partial_t \phi^{+}_{ai} ( A^+, B^+) C^-_i ~  
 \partial_t \phi^{-}_{ai} ( A^-, B^-) C^+_j  ) }
After rescalings of \largen\ the relations in 
\xiijxki\ allow us to show that 
polynomials in $ A^-,B^-$ can be pulled to the left 
of $C_i^-$ at the cost of converting the pair 
 $(A^-, B^-)$ to $(B^+, A^+)$ ( not the switch).
 There are extra terms coming 
from the RHS of \xiijxki\ which can be ignored in the 
leading large $n$ limit. The terms in \offkin\
can then be written as 
\eqn\offkini{ 
\int dt  ~ TR_{\cRnp}  ( \partial_t \phi^{+}_{ai} ( A^+, B^+) 
 \partial_t \phi^{-}_{aj} ( B^+, A^+)   C^-_i   C^+_j  ) }
It is useful to write this as a product of 
parts symmetric in $(ij)$ and a part antisymmetric 
in these indices. The symmetric part 
is simplified using \gs\ to replace 
\eqn\symcij{   ( C^-_iC^+_j  + C^-_jC^+_i) =  { 1 \over 2 } \delta_{ij}
              - A^{+}_{ki} B^{+}_{kj} -  A^{+}_{kj} B^{+}_{ki}  \equiv
G_{ij} }
The antisymmetric part is simplified using \txcom\ 
\eqn\asymcij{  ( C^-_iC^+_j  - C^-_jC^+_i)
= { - i \over 2 } \epsilon_{ijkl}( A^+_{kl} + B^+_{kl} ) 
\equiv H_{ij}  }
We have defined in \symcij\ and \asymcij\ 
a symmetric tensor and an  antisymmetric tensor
living on $S^2 \times S^2$. 
After converting the trace to an integral using 
$SO(4)$ invariance, 
we can write the kinetic terms in 
\offkin\ as 
\eqn\ofkin{ 
\int dt \int dA ~ dB ~  
\partial_t \phi^{+}_{ai} ( A, B) \partial_t \phi^{-}_{aj} ( B, A) 
( G_{ij} + H_{ij} ) }
Analogous to the term in \offkin\ there is 
a trace over $\cRnm$ which leads by similar steps as above to 
\eqn\ofkini{ 
\int dt \int dA ~ dB ~  
\partial_t \phi^{-}_{ai} ( A, B) \partial_t \phi^{+}_{aj} ( B, A) 
( G_{ij} - H_{ij} ) }
Although $H$ contains $i$ the action is hermitian since
hermitian conjugation converts $\phi^{+} $ to $ \phi^-$.

The above arguments illustrate the use of the algebraic relations 
of section 4 in deriving the action for fluctuations. 
 It appears that there is some $U(2) $ 
symmetry which mixes the $\cRnp$ and the $\cRnm$ 
but which is broken to $U(1) \times U(1)$. 
 We leave it to the future to elucidate all the symmetries
of such an action. A further interesting problem is 
to relate this theory on $S^2 \times S^2$ to some 
field theory on $S^3$. In the fuzzy even sphere case 
there was a $U(1)$ theory on the higher dimensional 
geometry and a non-abelian theory on the sphere 
itself \hdim. We would expect a generalization of that 
correspondence to yield some field theory on the sphere $S^3$. 
The counting of degrees of freedom in section 9 shows that 
 it cannot be a standard non-abelian theory. The discussion 
in section 9 will also show that the analogous field theory 
on the coset $ {SO(6) \over U(2) \times U(1)}$ might 
 be more easily related to a field theory on the five sphere.

\newsec{ The fuzzy five-sphere } 

\subsec{ Higher dimensional geometry from Stabilizer group } 

 A guess is that $End( \cRnp ) $  related to 
 ${SO(2k)\over U(k-1) \times U(1) }$. This works in the 
 case of $k=2$, and gives the right counting of degrees
 of freedom, i.e $N^2$ scales like $n^{10}$ while 
 quadratic expressions like $X_i^2$ scale like $n^2$. 
 This means we should expect a $10$ dimensional space 
 which is indeed the dimension of the above coset for 
 $k=3$.  
  Here we will develop an argument 
 similar to section 3.3 based on the expectation values of 
 generators of the matrix algebra evaluated in a state
 in $ \cRn $. We pick a state of the form  
 $ v_0 \otimes \cdots \otimes v_0 \otimes a_1^{\dagger} v_0 \cdots 
 a_1^{\dagger} v_0 $, where there are ${ (n+1) \over 2 } $ copies of 
 $v_0 $ and  $ {(n-1) \over 2} $ copies of $  a_1^{\dagger} v_0$. 
 We sum over different ways of embedding the copies  
 $ v_0$ and the $a_1^{\dagger} v_0 $   in $ V^{\otimes n }$
 in order to make sure we have a state in $ Sym ( V^{\otimes n } ) $.

Considering normalized quantities 
analogous to \largen, we have in the large $n$ limit
\eqn\expcts{\eqalign{
& <s| A_{12}^{+}   | s > = 1 \cr 
&   <s| A_{34}^{+}   | s > =  <s| A_{56}^{+}   | s > = 1 \cr
& < s | C_{1}^2 | s > = 1/2 \cr 
& < s | C_2^2 |s > = 1/2 \cr 
& <s| C_{1} C_{2} + C_{2} C_1 |s > = 0 \cr
& < s |  C_{1} C_{3} + C_{3} C_{1} |s >  = 0  ... \cr
& < s | B_{12}^+ |s > = -1 \cr 
&  <s| B_{34}^{+}|s>  = < s| B_{56}^{+}  |s>  =  1 \cr } }    
Other $A_{ij} $ and $B_{ij}$ have zero expectation value. 
The above variables are defined in the same 
way as for the fuzzy $3$-sphere.
These  expectation values 
in \expcts\ are only preseved by $ U(1) \times U(2) $. 
This means that the complete set of variables 
describing the matrix algebra $End ( \cRnp ) $ in the large $n$
limit have a solution which is stabilized by the 
subgroup $ U(1) \times U(2) $. The full set of solutions, 
obtained by action of $SO(6) $ on the irrep $ \cRnp $,  
is acted on transitively by the $SO(6) $. Hence the set of solutions 
is the coset $ SO(6)/U(2) \times U(1) $. 
The $U(1)$ is generated by rotations in the $12$ direction. 
In the $3546$  block there is a $U(2)$ subgroup 
which preserves the expectation values. 
Consider $ < X_{1}^2 >  = <X_2^2 > = 1 $. 
These are matrices of the form 
\eqn\stab{ \pmatrix{ P & Q \cr 
                    -Q & P \cr } } 
where $P$ is real antisymetric and $Q$ is real symmtric.

\subsec{Group Theory Proof : Harmonics of 
$SO(6)/U(2) \times U(1)$ from Matrix realization 
}

In the next two sections, we develop the 
arguments analogous to sections 3.1 and 3.2, 
now for the case of the fuzzy five-sphere. 
We recall from \sphdiv\ the list of representations 
of $SO(6)$ which appear in $End ( \cRnp )$ at large $n$. 
There are self-dual representations 
associated with highest 
  weights $ \vec \lambda = ( p_1 + p_2 + p_3 , p_1 + p_2 , p_1
)$, and antiself-dual reps with 
 highest  weights $ \vec \lambda = ( p_1 + p_2 + p_3 , p_1 + p_2 , - p_1
)$, where $p_1,p_2,p_3$ are positive. 
The associated Young diagrams have row lengths 
 $ \vec r = ( p_1 + p_2 + p_3 , p_1 + p_2 , p_1 )$
Unlike the case of the fuzzy three sphere, 
we now have representations with multiplicity 
more than $2$. 
The multiplicity is 
 zero unless the following conditions
are satisfied:   $p_1 + p_3 $ is even,  
 if $p_1$ is positive, $p_3 \ge p_1 $ and  
 if $p_1 $ is negative, $p_3 \ge |p_1| $. 
Then the multiplicity is 
\eqn\matmulpl{ m( p_1, p_2 , p_3 ) = p_2  + 1  } 
The multiplicity arises because of the multiple ways 
 of writing $p_2$ as a sum of two non-negative 
 integers $p_2^+$ and $p_2^-$. Different choices 
 of $p_2^+$ and $p_2^-$ lead to different operators 
 in $End ( \cRnp ) $ which transform according to 
 the same representation of $SO(6)$.

The space of functions on $ {SO(6 ) \over U(2) \times U(1)} $
is the $SO(6)$ representation induced from the
trivial rep. of $  U(2) \times U(1) $. 
By Frobenius duality, multiplicity of 
an $SO(6) $ irrep. in the induced rep is the same 
as the multiplicity of the trivial rep. of $U(2) \times U(1) $
in that $SO(6)$ representation \wal. 
 We will compare, therefore, the multiplicity 
 of the trivial rep. of $U(1) \times U(2)$
 in the restriction of any  given $SO(6)$ irreducible representation
 with the number of times that irrep. appears in 
 $ End( \cRnp)$.  We expect they are equal. 

 We saw in 8.2 that the multiplicities in $ End( \cRnp )$ are zero 
 unless $p_1 + p_3$ is even. This can be shown easily
 to be a property of the restriction multiplicities as well,  
 from facts about the root lattice of $SO(6)$ \fulhar. 
We show that unless $p_1 + p_3$ is an even integer, 
the multiplicity of appearance of 
the trivial rep. of $U(2) \times U(1) $ 
is zero.  A necessary condition for the trivial rep. to appear 
is that the multiplicity of the zero weight vector is non-zero. 
Unless $p_1 +p_3$ is even this multiplicity is actually zero. 

Indeed, the positive   roots are 
$ L_1 \pm  L_2 $,  $L_2 \pm L_3 $,
$L_1 \pm  L_3 $ \fulhar.    
The highest weight of the irrep. characterized by 
$ \vec \lambda  = ( p_1+p_2+p_3 , p_1 + p_2,  p_1 )$ 
is  $ (p_1 + p_2 +  p_3 ) (  L_1  ) + ( p_1 + p_2 ) ( L_1+  L_2   ) +
( p_1  ) (L_1 + L_2 + L_3 )  $. 
If any state of weight zero exists in the rep. 
it must be obtained by subtracting  from the highest weight
a positive root. So we need 
to write 
\eqn\exppr{ 
\vec \lambda = a_{1} ( L_1 + L_2  ) + a_2 ( L_1 - L_2 ) + a_3
( L_2 + L_3 ) + a_4 ( L_2 - L_3 )  + a_5 ( L_1 + L_3 ) + a_6 ( L_1 -
L_3 ) } 
If we add up the components of $\vec \lambda $ we get 
$ 2p_2 + 3 p_1 + p_3 $ which has to equal 
$ 2a_1 + 2 a_2 + 2a_3 + 2a_4 + 2 a_5 + 2 a_6$. 
Since the $a_i$ are positive integers, it is clear that this is only 
possible if 
$ p_1 + p_3 $ is even.

In the case where $p_1 + p_3$ is even 
but $ p_3 <  p_1 $ the multiplicity 
of the identity rep. of $U(1) \times U(2)$ 
is zero. The identity rep. is characterized 
by having a state of weight zero, which is annihilated
 by the raising or lowering operators of the 
$U(2)$. In the case $p_1 + p_3 $ odd, 
the multiplicity of the zero weight vector is 
zero, so it is clear that there is no trivial 
rep. In the case where $p_1 + p_3$ is even 
but $ p_3 <  p_1 $, the multiplicity 
of the zero weight vector is 
non-zero, but it belongs to a higher rep. 
of $U(1) \times U(2)$. 

We need to recall some details about the embedding 
of $U(1) \times U(2) $. Let $H_1, H_2, H_3 $
be the generators of the Cartan of $SO(6) $. 
$H_1 = L_{12}, H_2 = L_{34}, H_3 = L_{56} $. 
The $U(1)$ appearing as a factor in the subgroup 
is generated by $ H_1$. The abelian 
$U(1)$ of the $U(2)$ is generated by 
$H_2 - H_3 $. The Cartan of the $SU(2)$ subgroup 
of the $U(2)$ is $H_2 + H_3$. 
The trivial rep. of $ U(1) \times U(2)$ 
has $( H_1, H_2, H_3 ) = ( 0,0,0)$. 
Non-trivial reps. of this subgroup 
which also contain the zero vector 
have highest weights of the form $(0,k, -k  )$. 
Let $ m(0,k,-k) $ be the multiplicity of 
the state of weight $ (0,k,-k)$. 
Let $m_h ( 0,k,-k)$ be the multiplicity 
of $(0,k,-k)$ as a highest weight. 
For a fixed rep. of $SO(6)$ let $(0,K,-K)$ 
be the highest $U(1) \times U(2) $ weight which appears
in the restriction  to $U(1) \times U(2)$.   
The problem of finding the multiplicity of the trivial 
rep. is then reduced to calculating multiplicities
of such states. 
Indeed
\eqn\fndmlt{\eqalign{  
& m_h ( 0,0,0 ) = m (0,0,0 ) - m_h ( 0,0,0) - m_h ( 0,1,-1) 
 - \cdots m_h ( 0, K, -K ) \cr 
& m_h (0, 1, -1 ) = m (0,1,-1 ) - m_h ( 0,2,-2) - \cdots 
 m_h ( 0, K, -K ) \cr 
& m_h (0, 2, -2 ) =  m (0,2,-2 ) - m_h ( 0,3,-3) - \cdots  m_h ( 0, K,
-K ) \cr 
& \vdots \cr
& m_h ( 0,K, -K  ) = m ( 0, K, -K ) \cr }}
From these equations it follows 
that 
\eqn\mlttriv{ m_h (0,0,0 ) =  m (0,0,0 )   -  m (0,1,-1 ) } 

In the table below we give the relevant multiplicities
for some  of irreps. of $SO(6)$ and we calculate $m_h ( 0,0,0 )$. 
 The multiplicity 
obtained from the Matrix decomposition has been expressed
 in the final column using a $\theta $ function 
to express the fact that it is non-zero only when $ p_3  > |p_1| $
The multiplicities of the weight vectors are calculated 
on Maple using the Weyl character formula for $SO(6)$ 
 ( see for example \fulhar\ ).  We  observe 
that the multiplicity calculated using the 
character formula is equal to that we expect 
from the Matrix Decomposition, as shown 
by the agreement of the last two columns of the 
table. 

We have also performed the above checks when 
$p_1$ is negative, i.e the reps are anti-self-dual,
and we again get agreement between the Matrix decomposition 
multiplicities 
{}\vskip.5truein
\centerline{           
\begintable
\begintableformat
\center " \center " \center   " \center  " \center 
\endtableformat
\-    
\br{\:|}   $\vec \lambda $  |  $m( 0,0,0) $ | $ m(0,1,-1)$  | $m_h(0,0,0)$ |
                            $(p_2+1 ) \Theta(p_3-p_1)  $      \er{|}
\-
\br{\:|}    (2,0,0)  | 2 |    1 | 1 | 1      \er{|}
\-
\br{\:|}    (4,0,0) |    3 | 2 | 1 | 1     \er{|}
\-
\br{\:|}   (  6,0,0) |4  | 3| 1 | 1         \er{|}
\-
\br{\:|}    (8,0,0)  | 5   | 4 | 1 | 1         \er{|}
\-
\br{\:|}  (1,1, 0 ) | 3 | 1 | 2 | 2                              \er{|}
\-
\br{\:|}  (3,1, 0 ) | 7 | 5 | 2 | 2                              \er{|}
\-
\br{\:|}  (5,1, 0 ) | 11 | 9 | 2 | 2                              \er{|}
\-
\br{\:|}  (2,2, 0 ) | 6 | 3 | 3 | 3
\er{|}
\-
\br{\:|}  (4,2, 0 ) | 15 | 12 | 3 | 3
\er{|}
\-
\br{\:|}  (6,2, 0 ) | 24 | 21 | 3 | 3
\er{|}
\-
\br{\:|}  (3,3, 0 ) | 10 | 6 | 4 | 4
\er{|}
\-
\br{\:|}  (5,3, 0 ) | 26 | 22 | 4 | 4
\er{|}
\-
\br{\:|}  (7,3, 0 ) | 42 | 38 | 4 | 4
\er{|}
\-
\br{\:|}  (4,4, 0 ) | 15 | 10 | 5 | 5
\er{|}
\-
\br{\:|}  (6,4, 0 ) | 40 | 35 | 5 | 5
\er{|}
\-
\br{\:|}  (5,5, 0 ) | 21 | 15 | 6 | 6
\er{|}
\-
\br{\:|}  (2,1, 1 ) | 3 | 2 | 1 | 1
\er{|}
\-
\br{\:|}  (4,1, 1 ) | 6 | 5 | 1 | 1
\er{|}
\-
\br{\:|}  (6,1, 1 ) | 9 | 8 | 1 | 1
\er{|}
\-
\br{\:|}  (3,2, 1 ) | 8 | 6 | 2| 2
\er{|}
\-
\br{\:|}  (5,2, 1 ) | 16 | 14 | 2| 2
\er{|}
\-
\br{\:|}  (7,2, 1 ) | 24 | 22 | 2| 2
\er{|}
\-
\br{\:|}  (4,3, 1 ) | 15 | 12 | 3 | 3
\er{|}
\-
\br{\:|}  (6,3, 1 ) | 30 | 27 | 3 | 3
\er{|}
\-
\br{\:|}  (5,4, 1 ) | 24 | 20 | 4 | 4
\er{|}
\-
\br{\:|}  (4,2, 2 ) | 6 | 5  | 1 | 1
\er{|}
\-
\br{\:|}  (6,2, 2 ) |11 | 10  | 1 | 1
\er{|}
\-
\br{\:|}  (5,3, 2 ) |15 | 13  | 2 | 2
\er{|}
\-
\br{\:|}  (4,4, 2 ) | 6 | 6  | 0 | 0
\er{|}
\-
\br{\:|}  (4,3, 3 ) | 3 | 3  | 0 | 0
\er{|}
\-
\br{\:|}  (6,3, 3 ) | 10 | 9  | 1 | 1
\er{|}
\-
\br{\:|}  (7,4, 3 ) | 24  | 22  | 1 | 1
\er{|}
\-
\endtable }
\noindent
and those obtained from 
counting weights. It would be interesting to develop this 
proof using characters into an analytic proof but 
we will leave that to the enthusiastic reader, and hope 
the remaining readers will be content with the agreement 
demonstrated in the table below.

\subsec{ $Hom(\cRnp , \cRnm)$ as sections } 
 
We have matched the matrices which map 
$\cRnp$ to $\cRnp $, i.e endomorphisms \hfill\break
 $ End( \cRnp, \cRnp)$ 
to functions on the coset $ { SO(6) \over U(2) \times U(1) } $. 
The matrices  $ End( \cRnm, \cRnm) $ have the same decomposition 
into $SO(6)$ representations as  $ End( \cRnp, \cRnp)$. 
The matrices which map    $\cRnp$ to $\cRm $, 
or homomorphisms $ Hom ( \cRnp, \cRnm) $ have a 
different $SO(6)$ decomposition.
Here we will show that  the 
$SO(6)$ decomposition 
of $Hom( \cRnp, \cRnm)$ matches that of sections 
of a certain bundle over  $ { SO(6) \over U(2) \times U(1) } $.
 
Sections of bundles over a coset $G/H$ are 
induced representations.
They are specified by choosing a representation $R_H$ 
of the subgroup $H$ constructing the representation of 
$G$ induced from it.  In the discussion above, $R_H$ was the
trivial bundle. 
 By the Frobenius reciprocity 
theorem, the multiplicity of an irreducible  representation 
of  $R_G$ of $G$  in the induced representation is equal to 
the multiplicity of $R_H$ in the restriction 
of $R_G$ to $H$.
Sections of a non-trivial bundle are obtained by chooosing a
non-trivial representation of $H$. 
Here we show that $Hom( \cRnp, \cRnm)$ matches 
the representation of $SO(6) $ induced from 
$ (1, 0, 0 )  $ of $U(1) \times U(2) $. The first 
integer here is the $U(1) $ charge and the last two integers
label the representation of $U(2)$. 
 $Hom( \cRnm, \cRnp)$ matches 
the representation of $SO(6) $ induced from 
$ (- 1, 0, 0 )  $ of $U(1) \times U(2) $.

We will give  the evidence in the case 
of  $Hom( \cRnp, \cRnm)$. 
Let us recall from \sphdiv\ the multiplicities 
of irreps. of $SO(6)$ appearing in  $Hom( \cRnp, \cRnm)$. 
For positive integers $p_1,p_2,p_3$  there  are self-dual 
representations $( \lambda_1, \lambda_2, \lambda_3 )  
= ( p_1 + p_2 + p_3, p_1 +p_2, p_1 ) $. 
The multiplicities are $ m ( \lambda ) = p_2 + 1 $ 
if $ p_1 + p_3 $ is odd and $p_3 \ge  p_1 - 1 $ 
If these conditions on $p_1 $ and $p_3 $ are not satisfied, 
then the multiplicity is zero. At finite 
$n$ there is an  upper limit  $p_1 +p_2 +p_3 \le n $, 
but as $n \rightarrow \infty $ there is no upper limit. 
There are also anti-selfdual representations 
labelled by $( \lambda_1, \lambda_2, \lambda_3 )  
= ( p_1 + p_2 + p_3, p_1 +p_2, - p_1 ) $, 
so that $ \lambda_3 $ is negative, while $p_1,p_2,p_3$ 
are still positive. Again the condition for 
a non-zero multiplicity is that $ p_1 + p_3 $ is odd, 
and that $p_3 \ge p_1 + 1   $. When these conditions are satisfied, 
the multiplicity is given by $ m( \vec \lambda )  = p_2 + 1 $.  
These multiplicities from the Matrix deomposition
have to agree with restriction multiplicities, i.e 
the number of times $(1,0,0)$ appears 
as a $U(2) \times U(1) $ highest weight 
for any given   highest weight $\vec \lambda $  
of $SO(6)$. 

The condition $ p_1 + p_3 $ be odd is easy  to 
prove ( just as the evenness of $p_1+p_3$ is easy 
to prove in the case of restriction multiplicities 
 for the trivial rep. of $SO(6)$).  
It simply follows from the 
fact that if $p_1 +p_3$ is even, 
the corresponding highest weight differs
from the vector  $(1,0,0)$ by an element which is not 
in the root lattice. This means that the multiplicity 
of $(1,0,0)$ is zero. In the case where $ p_1 +p_3$ 
is odd but the inequalities between $p_1$ and $p_3$ 
are not respected, the multiplicity of 
the weight vector $(1,0,0)$ is non-zero, 
but its multiplicity as a highest weight of 
$U(1) \times U(2) $ is zero.

{}\vskip.2truein
\centerline{  
\begintable
\begintableformat
\center " \center " \center   " \center  " \center 
\endtableformat
\-    
\br{\:|}   $\vec \lambda $  |  $m( 1,0,0) $ | $ m(1 ,1,-1)$  | $m_h(1,0,0)$ |
                            $(p_2+1 ) \Theta(p_3-p_1 + 1  )  $      \er{|}
\-
\br{\:|}    (1,0,0)  | 1 |    0 | 1 | 1      \er{|}
\-
\br{\:|}    (3,0,0) |    2 | 1 | 1 | 1     \er{|}
\-
\br{\:|}   (  5,0,0) |3  | 2| 1 | 1         \er{|}
\-
\br{\:|}    (7,0,0)  |  4  | 3 | 1 | 1         \er{|}
\-
\br{\:|}  (2,1, 0 ) | 4 | 2 | 2 | 2                              \er{|}
\-
\br{\:|}  (4,1, 0 ) | 8 | 6 | 2 | 2                              \er{|}
\-
\br{\:|}  (6 ,1, 0 ) | 12 | 10 | 2 | 2                              \er{|}
\-
\br{\:|}  (8 ,1, 0 ) | 16   |  14  | 2 | 2                              \er{|}
\-
\br{\:|}  (3,2, 0 ) | 9 | 6 | 3 | 3
\er{|}
\-
\br{\:|}  (5,2, 0 ) | 18 | 15 | 3 | 3
\er{|}
\-
\br{\:|}  (7,2, 0 ) | 27| 24 | 3 | 3
\er{|}
\-
\br{\:|}  (4,3, 0 ) | 16  | 12 | 4 | 4
\er{|}
\-
\br{\:|}  (6,3, 0 ) | 32 | 28 | 4 | 4
\er{|}
\-
\br{\:|}  (5,4, 0 ) | 25 | 20 | 5 | 5 
\er{|}
\-
\br{\:|}  (3,1, 1 ) | 4 | 3 |1  | 1
\er{|}
\-
\br{\:|}  (5,1, 1 ) | 7 | 6 | 1 | 1
\er{|}
\-
\br{\:|}  (7,1, 1 ) | 10 | 9 | 1 | 1
\er{|}
\-
\br{\:|}  (5,2, 2 ) | 8 | 7 | 1 | 1
\er{|}
\-
\br{\:|}  (7,2, 2 ) | 13 | 12 | 1 | 1
\er{|}
\-
\br{\:|}  (4,3, 2 ) | 8 | 6 | 2 | 2
\er{|}
\-
\br{\:|}  (6,3, 2 ) | 20 | 18 | 2 | 2
\er{|}
\-
\br{\:|}  (3,3, 3 ) | 1 | 1 | 0 | 0
\er{|}
\-
\br{\:|}  (5,3, 3 ) | 6 | 5 | 1| 1
\er{|}
\-
\br{\:|}  (5,4, 4 ) | 3 | 3| 0| 0
\er{|}
\-
\br{\:|}  (7,4, 4 ) | 10 | 9 | 1| 1
\er{|}
\-
\endtable }
\noindent
The multiplicity of 
$ (1,0,0)$ as a highest weight of $ U(1) \times U(2) $,
which we denote as $m_h( 1,0,0)$, 
is related to the multiplicities
of weights of the form $ (1, 0,0) ,~~ ( 1,1,-1) ,~~ \cdots $ \hfill\break
$(1,K,-K )$ where $(1,K,-K )$ is the highest weight 
of this form appearing in the irrep. specified by the highest 
weight $ \vec \lambda $. 
 $ m( 1,0,0)$ and 
$ m(1,1,-1) $ by 
\eqn\mhmb{\eqalign{ 
&  m_h( 1,0,0) = m( 1,0,0) -  ( m_h  (1,1,-1) + \cdots + m_h ( 1, K,-K
)    ) \cr 
&  m_h ( 1,1,-1 ) = m (1,1,-1) - ( m_h ( 1,1,-1) + \cdots  m_h ( 1,
K,-K )) \cr 
& \vdots \cr 
&  m_h ( 1,2,-2) = m(1, 2,-2) -  ( m_h ( 1, 2 , -2 ) + \cdots m ( 1,
K, -K ) )   \cr 
& m_h (1, K, -K ) = m ( 1, K , -K ) \cr }} 
which leads to a simple equation for 
$ m_h( 1,0,0) $ in terms of $m( 1,0,0) $ and $ m(1,1,-1)$. 
\eqn\mhmbsmp{  m_h( 1,0,0) = m( 1,0,0) - m (1,1,-1) }

The table above gives a counting of weight multiplicities 
performed with Maple using the Weyl character formula
for $SO(6) $ and shows that the counting of
multiplicities of the $U(2) \times U(1) $ irrep.
obtained by restriction of $SO(6)$ agrees with 
that obtained from analysing the $SO(6)$ content 
of $ Hom ( \cRnp. \cRnm ) $.

\newsec{ Some remarks on geometry and combinatorics of fuzzy odd
spheres } 

\subsec{ geometry of fuzzy $S^{2k}$ } 
 
 The coset ${ SO(2k) \over U(k -1  ) \times U(1) } $
can be seen to  admit the following bundle structure,  
\eqn\abndtk{\eqalign{  
& { SO(2k) \over U(k-1) \times U(1) }  \longleftarrow 
{ U(k)  \over U(k-1) \times U(1) }   \cr 
& \qquad  \downarrow \cr 
&  { SO(2k) \over U(k) } \cr }}
using the general bundle structure 
$H/K \rightarrow  G/K \rightarrow G/H$
which exists whwnever we have subgroups $ K \subset H \subset G$. 
The base is a hermitian symmetric space ( incidentally  
 one of the kind that comes up as the fibre of 
 the bundle structure exploited in \hdim\ for 
 fuzzy even spheres). The fibre here is actually a 
  coset description of $CP^{k-1}$, also a symmetric space. 
 In the case
 $k=2$, the base and fibre are both $S^2$. 
In the case $k=3$, the base is $CP^3$ and 
the fibre $CP^2$. In the case $k=4$, the fibre is $CP^4$ 
 and the base is $SO(8)/U(4)$. 

In the case $k=2$, the base $S^2$ is of lower dimension 
than the sphere $S^3$ we started with. For higher 
$k$ the base of the above fibration is always 
of higher dimension than the odd sphere $S^{2k-1}$. 
The case $k=3$ of the fuzzy five sphere is 
 rather special in that the $CP^3$ is simply related to 
 the $S^5$ by a $U(1)$ quotient, $ S^5 = CP^3/U(1)$, 
 a fact which found a recent physical application 
 in \atwit\ for example. In \hdim\ the  bundle structure 
 of the higher dimensional geometry was used to 
 obtain a non-abelian theory on the sphere. It is 
 possible that the special relation of $CP^3$ to $S^5$ 
 will allow the Matrix algebra related to fuzzy $5$-sphere 
 to be related to field theory on 5-sphere, and hence 
to brane physics of spherical branes, more easily 
 than the other cases. It is intriguing that the physics 
of type IIA theory which contains both supersymmetric 
 zero-branes and NS-5 branes might also be taken to suggest, 
 using a simple-minded asumption that zero-branes should 
 in some context blow-up into NS-5,  
 that fuzzy 5-spheres would be easier than other odd 
 spheres.

Another possibly useful bundle structure is 
\eqn\anbnd{\eqalign{  
& { SO(2k) \over U(k-1) \times U(1) }  \longleftarrow 
{ SO(2k-2) \times U(1)  \over U(k-1) \times U(1) }   =
{ SO(2k-2)  \over U(k-1) }  \cr 
& \qquad  \downarrow \cr 
&  { SO(2k) \over U(1) \times SO(2k-2)} \cr }}

\subsec{ Projecting to $S^{2k-1}$ and possible role for $S^{2k-1}/Z_2
=  RP^{2k-1} $ } 

 In the case of the fuzzy even sphere, 
 there is a simple way to project 
 the Matrix algebra onto the $SO(2k+1)$ reps 
 which are symmetric traceless, and which  
 approach the algebra of functions on a 
 sphere at large $n$. This gives rise 
 to a non-associative algebra at finite $n$ \sphdiv\hdim. 
 We can apply a projection to get a 
 space of functions which is the same
 as the space of functions on $S^{2k-1}$. 
 This requires the off diagonal components. 
 If  we only keep  Matrices which transform as 
 symmetric traceless irreps. of $SO(2k-1)$ and 
 and which are   invariant under conjugation by a permutation which 
 switches positive chirality  with negative, this will
 give us the desired representations.  
 This projects out the  $Y_i$, while 
 keeping $X_i$ and their symmetric products. 
 It also kills off $X^{\pm}_{ij}, Y^{\pm}_{ij} $ 

 If we look at the upper block alone  $ End ( \cRnp) $ and 
 project out antisymmetric representations 
 we are left with the algebra of functions on $RP^{2k-1} = S^{2k-1}/Z_2$. 
  The matrices transforming as a vector, i.e the $X_i$, 
 are projected out as well as those which 
 transforming according to any $SO(2k) $ Young diagram with 
 $\vec r = (r_1,0 \cdots ) $ and $r_1$ odd.   
 The advantage of working with $RP^{2k-1}$ is that in the large 
 $n$ limit, the algebra becomes associative as in the case 
 of the algebra of the fuzzy four-sphere and for similar 
 reasons. The correlation between spatial $SO(2k) $ transformation 
 properties and a $Z_2$ subgroup of a $U(2)$ is reminiscent 
 of structures encountered in the study of zero branes 
 on compact orbifolds in \ramwal\grlayi. These analogies
 may be useful in finding physical brane realizations
 of fuzzy odd spheres.   

\subsec{Combinatoric  comparison of even and odd spheres }

 In the even fuzzy sphere   cases we can deduce 
the rank of the non-abelian group by 
looking at the geometry. The radius always scales 
as $n$. The number of degrees of freedom ( dof )  required to 
describe the sphere scales as $n^{2k}$ for $S^{2k}$. 
The total number of dof  scales as $N^2$, where 
$N$ is the size of the matrix, i.e the dimension
of $\cRn$. It is also equal to $n^{D} $ where 
$D$ is the dimension of the higher dimensional coset. 
When we formulate the physics as a non-abelian 
one, the rank is $n^{D - 2k\over 2  }$.  This is
made concrete in the table below.

{}\vskip.5truein
\centerline{           
\begintable
\begintableformat
\center " \center " \center   " \center  " \center " \center 
\endtableformat
\-    
\br{\:|}   Fuzzy sphere   | $n$-scaling of   | $n$-scaling of | 
$n$-scaling  of 
| n-scaling of  | $n$-scaling for  rank \er{|} 
\br{\:|}  $S^D$ |  Radius  | geometrical | 
 total dof  $N^2
\sim  n^T $ | excess dof |
  of gauge \er{|}
\br{\:|} $D=2k $   | $R \sim n $ |  dof ~~~   $n^D$ |  
 $ = ~~ n^{ k^2 + k }   $ | $n^{T-D} $  | 
 group $ = n^{T-D\over 2}$   \er{|}
\-
\br{\:|}   $S^2$   | $n$ |    $n^2$ | $n^2$ | $1$ | $1$      \er{|}
\-
\br{\:|}  $S^4$ | $n$ | $n^4$  | $n^6$ | $n^2$ | $n$ \er{|}
\-
\br{\:|}  $S^6$  | $n$ | $n^6$  | $n^{12}$ | $n^6$ | $n^3$ \er{|}
\-
\br{\:|}  $S^8$ | $n$ | $n^8$  | $n^{20}$ | $n^{12}$ | $n^6$  \er{|} 
\-
\endtable }
\medskip

It is instructive to compare this with fuzzy 
odd spheres

{}\vskip.5truein
\centerline{           
\begintable
\begintableformat
\center " \center " \center   " \center  " \center 
\endtableformat
\-    
\br{\:|}   Fuzzy sphere   | $n$-scaling of   | $n$-scaling of | 
$n$-scaling  of 
| n-scaling of \er{|}
\br{\:|}  $S^D$ |  Radius  | geometrical | 
 total dof  $N^2
\sim  n^T $ | excess dof \er{|}
\br{\:|} $D=2k-1 $   | $R \sim n $ |  dof ~~~   $n^D$ |  
 $ = ~~ n^{ k^2+ k-2 } $ | $n^{T-D} $  \er{|}
\-
\br{\:|}   $S^3$   | $n$ |    $n^3$ | $n^4$ | $n$    \er{|}
\-
\br{\:|}  $S^5$ | $n$ | $n^5$  | $ n^{10} $ | $n^5$    \er{|}
\-
\br{\:|}  $S^7$  | $n$ | $n^{7}$  | $n^{18}$ | $ n^{11} $   \er{|}
\-
\br{\:|}  $S^9$ | $n$ | $n^{9}$  | $n^{28}$ | $n^{19} $ \er{|} 
\-
\endtable }

The first five columns for even and odd spheres 
are the same, and the odd sphere analog of the last column 
is conspicuously absent, since we have not 
been able to decipher the nature of the gauge theory 
living on the odd sphere, although we did give an outline 
 of a field theory on the higher dimensional coset in section 7. 
Since the number of extra degrees of freedom
scales like an odd power of $n$, assuming one could
 make sense of the theory on the odd sphere as 
 some sort of generalized non-abelian gauge theory, 
 the rank would be $n^{p/2}$ for $p$ odd. 
 Curiously, a completely different line of argument, 
 also motivated by an attempt to understand 
5-branes, leads one to consider  a generalization of non-abelian 
 gauge theory where the number of degrees of freedom 
 scales like $k^3$ rather than $k^2$ where $k$ is 
 an integer number of branes. This famous exponent 
 appears in trying to account for the entropy 
 of five-branes \klebtsey.  The exponent $n^5$ should be compared 
 to the odd power of $k$ \foot{ I thank Savdeep Sethi 
 for discussions along these lines }.  It would be interesting to find
a scenario where the appearance of odd powers of a brane 
number as the counting of degrees of freedom of a field 
 theory on a fuzzy odd sphere can be related to the entropy of 
five branes.

\subsec{ Further comments and open questions } 

The results of \clt\sphdiv\myrsnabint\hdim\ 
show that a lot of interesting physics 
of spherical $4$-branes constructed 
from $0$ branes  can be inferred 
directly from  the Matrices $X_i$ satisfying 
the defining equation of an even  sphere. 
It is also clear that the higher dimensional 
fuzzy spheres give a neat generalization 
of the physics of fuzzy $4$-spheres \hdim\fab. 
In this paper, we hope to have convinced the reader
that by considering a natural generalization of
the fuzzy even sphere construction to the odd sphere
case as in \guram\ we can find many features 
similar to the fuzzy even sphere case, e.g 
the relevance of field theories on higher dimensional 
cosets. Admittedly these field theories are somewhat more 
complicated and even the sketch of a reduction to a theory on the 
sphere itself has not been accomplished ( unlike the even sphere 
case  ). However the fact that the higher dimensional 
cosets have a rather rich geometry related to symmetric spaces, 
as explained in section 9.1 suggests that these fuzzy odd 
sphere Matrix constructions should have a role similar in 
at least some respects to fuzzy even spheres. 
One question in the direction of finding this physics 
is to find a SUSY Matrix quantum mechanics model, 
perhaps allowing the generalized type of SUSY  
considered in \bmn, which would admit any of 
the fuzzy odd spheres as a SUSY preserving 
solution.
It was observed in \bmn, using simple 
scaling arguments, that one did not expect 
a five sphere to appear as a classical solution. 
It would be interesting to find {\it any } M-theory  background 
with a Matrix model dual  where the duality 
would predict the existence, rather the non-existence, 
of fuzzy five spheres. This may be viewed as a reformulation 
of the transverse $5$-brane problem of BFSS Matrix theory.

\vskip2.0in

\noindent{\bf Acknowledgements } 

\vskip.2in
I am happy to acknowledge interesting discussions with  Bobby Acharya, 
Steve Corley, Jerome Gauntlett, Zack Guralnik,  George Papadopoulos,
Chris Hull, Antal Jevicki, Robert de Mello Koch, Rob Myers,   Savdeep Sethi, 
Wati Taylor. This research  was supported
by DOE grant  DE-FG02/19ER40688-(Task A). I would like to thank the
Isaac Newton Institute for hospitality while part of this work was
done.

\vskip0.5in

\noindent{\bf  Appendix 1 : Notation for Gamma matrices}

We recall the explicit construction of the $\Gamma$ matrices. 
 To fix notation we will give the form of the 
 Gamma matrices $\Gamma_{\mu} $ with $\mu=1\cdots 2k+1$. 
The Gamma Matrices of $SO(2k+1)$  obey the equations :  
\eqn\gam{ 
[\Gamma_{\mu} , \Gamma_{\nu} ] = \delta_{ \mu \nu }  } 
They can be expressed 
 in terms of a set of fermionic  oscillators $a_{i}$ with 
$i$ running from $1$ to $k$, and obeying 
\eqn\fermdef{ \{ a_{i}, a_{j}^{\dagger} \} = \delta_{ij} }. 
The expressions for the Gamma Matrices are :  
\eqn\gamrep{\eqalign{  
& \Gamma_{2i-1} =  ( a_{i} + a_{i}^{\dagger} )   \cr 
& \Gamma_{2i} = i ( a_{i} - a_{i}^{\dagger}   ) \cr }}
 for $ i = 1 \cdots k $  and   
$ \Gamma_{2k+1} = i^{k} \Gamma^{1} \cdots \Gamma^{2k} $. 
 From \gamrep\ it follows that 
$\Gamma_{2i-1} \Gamma_{2i}   = i [a_i^{\dagger}, a_{i} ] $. 
 
 A $2^k$ dimensional representation of the 
 $ \Gamma $ matrix algebra is obtained by defining 
 a state $ v_0 $ which is annihilated by the 
 fermionic annihilation operators $a_{i}$ and
 acting with the creation operators to generate  
 $2^{k} $ different states.  
 Using the abobe we find $ \Gamma^5 v_0 = v_0 $.
 States in $V_+$, the positive  
 chirality representation of $SO(2k)$, 
 have an even number of $ a^{\dagger}$ 
 acting on $v_0$. States in $V_-$, the negative 
 chirality representation of $SO(2k)$,  have an odd number 
 of $a^{\dagger}$ acting on $v_0$.

\bigskip

\listrefs

\end